\documentclass[a4paper,10pt]{article}
\usepackage[T1]{fontenc}
\usepackage[utf8x]{inputenc}
\usepackage[italian,english]{babel}
\usepackage[autostyle,italian=guillemets]{csquotes}
\usepackage{multicol}
\usepackage{microtype}
\usepackage{guit}
\usepackage{booktabs}
\usepackage{graphicx}
\usepackage{mathptmx}
\usepackage{amsfonts}
\usepackage{amssymb}
\usepackage{amsmath}
\usepackage{ifpdf}
\usepackage{dcolumn}
\usepackage{booktabs, longtable}
\usepackage{hyperref}
\usepackage{textcomp}
\usepackage{tabularx,array}
\usepackage{ulem}
\usepackage{placeins}
\usepackage{epstopdf}
\usepackage{etex}
\usepackage{rotating}
\date{}
\title{New Astronomical, Meteorological and Geological Study of Montefiascone (VT)} 
\author{Tasselli, D. \\ TS Corporation Srl - Astronomy and Astrophysics Department\\ Via Rugantino, 71 - 00169 Roma RM - Italy \\ E-mail:diego.tasselli@tscorporation.org \\ \\ Ricci, S. \\ TS Corporation Srl - Meteorogical and Climatic Change Department\\ Via Rugantino, 71 - 00169 Roma RM - Italy \\ E-mail:stefano.ricci@tscorporation.org \\ \\ Bianchi, P. \\ TS Corporation Srl - Geology and Geophysics Department\\ Via Rugantino, 71 - 00169 Roma RM - Italy \\ E-mail:pamela.bianchi@tscorporation.org}
\date{}
\begin{document}
 
\maketitle
 
\begin{abstract} 
\normalsize In this work that continues the "NGICS - New Italian City Geological Study" a project of the Department of Climatology and Geology of TS Corporation Srl, we present the study relating to the Municipality of Montefiascone (VT). We analyzed 25 years of astronomical, geological, meteorological and climatic data, comparing them to verify the long-term trend of local variations in temperatures, detections, solar radiation and geological events, with the ultimate goal of understanding climate and geological changes a long term in this geographical area. The analysis is performed using a statistical approach and attention is used to minimize any effects caused by the error in case of lack of data. 
\end{abstract}
\footnotesize{This paper was prepared with the \LaTeX} \\
\begin{multicols}%
{2}
\section{\normalsize Introduction} \normalsize In this work we present an analysis of the astronomical, geological, meteorological and climatic data collected in Montefiascone (VT), a small town in northern Lazio, located on the southern shore of Lake Bosena.\\We present an analysis of measurements obtained from local astronomical, meteorological and geological data \cite {datimeteo:ref2013} \cite {eumetsat:ref2013} \cite{Barberi:2013}, compared to verify local variations in astronomical conditions, climatological and geological. \\
\subsection{\normalsize Location} \normalsize Montefiascone is a small town located north of the Tuscan region in Italy. Here are identifiable altimeter data, the geographic coordinates and seismic data.
%Inserire tabella con coordinate  ed altezza
\begin{tabular}{c|c|c}
\hline
\scriptsize \textbf{Latitudine} & \scriptsize \textbf{Longitude} & \scriptsize \textbf{Altitude mt s.l.} \cr
\hline
\scriptsize $42^\circ 32' 25,08$'' N & \scriptsize $12^\circ 2' 13,56$'' E & \scriptsize 590 \cr
\hline 
\end{tabular}  \\  \\ \scriptsize {\bf {Geographic Data of Montefiascone}} \\ \\
\normalsize The geological map of the town is shown in Geological and Seismic data section. \cite{mappageologica:2013} \\
%Inserire tabella con coordinate  ed altezza
\begin{tabular}{lp{0.2\textwidth}}
\hline
\scriptsize \textbf{Seismic Zone} & \scriptsize \textbf{Description}  \cr
\hline 
\scriptsize 2B & \scriptsize Area with medium seismic hazard where strong earthquakes can occur. Sub-area 2B indicates a value of ag <0.20g. \cr
\hline 
\end{tabular} \\ \\ \scriptsize {\bf {Seismic Data of Motefiascone}}
\normalsize \section{\normalsize Annual data analysis} \normalsize Summers are generally hot, dry and the sky is clear or partly cloudy, while winters are rather harsh and often blow north winds that can last a long time especially in the period between November and December. In winter mornings it is easy to find fog also due to the proximity of Lake Bolsena. Snow appears about two or three times a year, sometimes in large quantities. Rains dominate the autumn and spring seasons.\\ \normalsize The following table identifies the climatic data assigned by the President of the Republic's Decree n. 412 of 26 August 1993.\\ Characterized by a fairly humid climate, the summer does not record maximum temperatures higher than the area thanks to the mitigation brought about by Lake Bolsena, while in winter there can be substantial variations in temperatures, due to the winds that blow more in the period between November and January. \\ \\
%Inserire tabella con coordinate  ed altezza
\begin{tabular}{c|c}
\hline
\scriptsize \textbf{Climatic Zone} & \scriptsize \textbf{Day Degrees} \cr
\hline 
\scriptsize E & \scriptsize 2.467 \cr
\hline 
\end{tabular} \\ \\ \scriptsize {\bf {Climatic Parameter of Montefiascone}} \normalsize \section{\normalsize Meteo-Climatic Parameter} \normalsize In this section we describe air temperatures (T), Dew Point, Humidity, Pressure, Day Time and Night Time Variation, Rain's Days and Fog's Day, obtained by an accurate analysis of the meteorological data from local data by archive \cite{datimeteo:2013}. We analyzed the parameters given in Table 1 and 2. \\Should be noted that the values considered are related to the last twenty-four-year average and made available for the period 1990-2014. \\ \\
\begin{tabular}{lp{0.10\textwidth}} 
\hline
{\footnotesize Average Annual Temperature}&  {\footnotesize $14,40 ^\circ C$}  \cr
{\footnotesize T average warmest (ago-2003)}& {\footnotesize $39,10 ^\circ C$} \cr
{\footnotesize T average coldest (feb-1991)}&  {\footnotesize $-10,00 ^\circ C$} \cr 
{\footnotesize Annual temperature range}&  {\footnotesize $9,87 ^\circ C$}  \cr
{\footnotesize Months with average T > $20 ^\circ C$} & {\footnotesize  199} \cr
{\footnotesize Total rainfall 1990-2014 [mm]}& {\footnotesize 37586,20} \cr
{\footnotesize Rain Days }&  {\footnotesize 1281}  \cr
{\footnotesize Fog Days }&  {\footnotesize 923}  \cr
{\footnotesize Storm Days} &  {\footnotesize 137}  \cr
{\footnotesize Rain/Storm Days }& {\footnotesize 490}  \cr
{\footnotesize Rain/Snow Days}&  {\footnotesize 26}  \cr
{\footnotesize Rain/Fog Days} & {\footnotesize 136}  \cr
{\footnotesize Rain/Thunder/Fog Days} & {\footnotesize 66}  \cr
{\footnotesize Snow Days} & {\footnotesize 36}  \cr
{\footnotesize Wind Speed max Km/h (Mar-2001)} & {\footnotesize 100,00}  \cr
{\footnotesize Wind Speed min Km/h (Mar-1990)} & {\footnotesize 1,00}  \cr
{\footnotesize Rain max mm (Aug-1995)} & {\footnotesize 660,00}  \cr
{\footnotesize Rain min mm (May-2006)} & {\footnotesize 2,50}  \cr
{\footnotesize Earthquake Min (2011/12/20)} & {\footnotesize 0,2 Mw}  \cr
{\footnotesize Earthquake Max (1992/07/02)} & {\footnotesize 3,6 Mw}  \cr
{\footnotesize Earthquake Deep Min (2008/06/03)} & {\footnotesize 0,8 Km}  \cr
{\footnotesize Earthquake Deep Max (2010/05/18)} & {\footnotesize 37,3 Km}  \cr
\hline
\end{tabular} 
\\ \\ \mbox{\bf{\footnotesize  Parameter of this Study}}
\subsection{Solar Radiation Territory} \normalsize In the study area the values of solar radioation are increasing. In fact the data irradiation of territory taken from the parameters and the data prepared by the European Union, demonstrate the trend of irradiation for the municipality, visible in next table: \\
% Table generated by Excel2LaTeX from sheet 'Calcolo della radiazione solare'
\begin{tabular}{|c|c|c|}
\hline
\textbf{Month} & \multicolumn{1}{c}{\textbf{DNI}} & \multicolumn{1}{c|}{\textbf{Hh}} \\
\hline
Jan   & 3698,40 & 1905,10 \\ 
Feb   & 4259,00 & 2654,60 \\ 
Mar   & 5544,60 & 3895,30 \\ 
Apr   & 6227,20 & 4860,50 \\ 
May   & 7634,20 & 6091,40 \\ 
Jun   & 8098,30 & 6528,00 \\ 
Jul   & 8152,80 & 6501,70 \\ 
Aug   & 7125,50 & 5580,20 \\ 
Sep   & 5826,70 & 4324,20 \\ 
Oct   & 4606,70 & 3062,90 \\ 
Nov   & 3707,40 & 2088,20 \\ 
Dec   & 2953,20 & 1533,50 \\ \hline
\end{tabular}%
\\ \\
\mbox{\bf{\footnotesize DNI - Direct Normal Irradiance and Radiation }} 
\mbox{\bf{\footnotesize and Hh - Radiation on Orizontal Plan (Wh/m$^2$/day) }}\\ \\
\normalsize The weather data and the graphs show extrapolated for the territory covered by the study, including a radiation in the range between 1100 and 1200 kWh /1kWp as map prepared by the European Union \cite{UE:2013} and visible in figure 17, characterized in over the months, irradiation presented in the graph in Figure 18 and 19, which shows the data of the table above, which shows the territory of Montefiascone, a total irradiance Annual of 2067,28 Wh/m$^2$/day. 
\subsection{Temperature} \normalsize {In this section we describe air temperatures (T) obtained by an accurate analysis of the meteorological data. The study shows a non-constant average in the temperatures recorded in the period between 1990 and 2014, alternating cold and hot periods, as shown in the following graph.} \\
\includegraphics[width=0.49\textwidth{}]{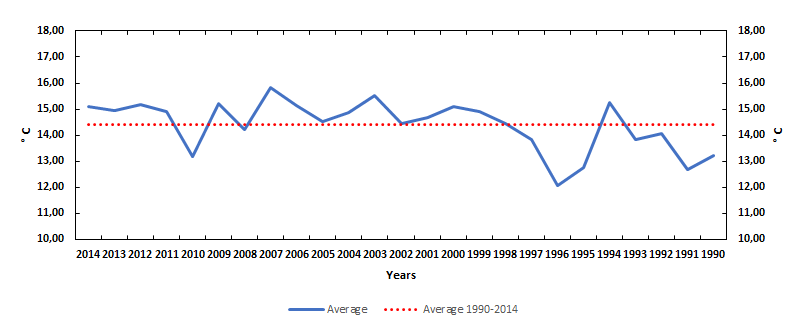} 
{\scriptsize \bf \textit{\textit{{\footnotesize Relationschip of Year Temperature and Average 1990-2014}}}}. \\
 \\The values are calculated considering the entire measurement period (1990-2014), drawing on data from the Annuals published by "Il Meteo.it" \cite{datimeteo:2013} for the period 1990-2014. \\ A comparative analysis in the years 1990-2003 shows that the average monthly temperatures were around the average, reaching the maximum peak in August and the lowest peak in December. A greater deviation of the values is highlighted in the periods ranging from January to March, from June to August and from October to December, while the period between the beginning of March and mid-April remains stable on the annual average. \\ \\ \includegraphics[width=0.50\textwidth{}]{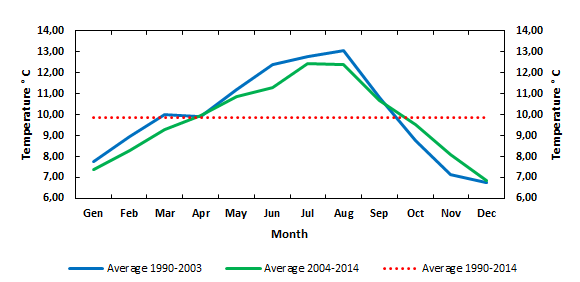} 
 {\scriptsize \bf \textit{\textit{{\footnotesize Variation of Month Temperature about year 1990-2003 and 2004-2014}}}} \\ \\
 The graphs shown in Figures 2 and 3 show the performance of the maximum and minimum temperatures proportional to the average of the period of study (1990-2014). The study shows the following trends:
\begin{itemize}
\item Minimum temperatures: recorded significant variations in the values during the period, always constantly rising, except in five years (1991, 1993, 1996, 2005 and 2010), in which they returned strongly below the average, which is equal to $3,94^\circ C$;
\item Maximum temperatures: The maximum temperatures also had a further increase, more evident in the period after 1996. In fact, the analysis of the data showed that if in the period between 1990 and 1999, the values were constantly below the average (except in the period between mid-1993 and early 1994), starting from the 2000s temperatures have been constantly increasing, except for a short period of reduction between 2008 and 2010), to return to average values in 2014, which is equal to $24.86^\circ C$.
\end{itemize} 
Table 6 shows the trend of the maximum temperature points forecast in the summer quarter (June-July-August) shown in Figure 13 and the trend of the minimum temperature points forecast in the winter quarter (January-February-March) and are shown in Figure 14. Studying in particular the months with the values of minimum temperature and maximum minors, the following is noted:
\begin{itemize}
\item the month of August (characterized by the presence of the highest value observed in maximum temperatures), shows that trends in temperature has been getting consistently below average for the period 1990 to 2014, while there were two exceedances of this value over the years 2003 and 2012, in agreement with what evident from the graph in Figure 12;
\item the month of February (characterization from the the lowest minimum temperature for the period of study), shows that the trend of temperatures has always been, in agreement with what reported in the graph in Figure 13.
\end{itemize}
\subsubsection{Average temperature fluctuation 1990-2014}
Analyzing the temperature data in the period 1990-2014, it can be seen as shown in table no. 8 and from the following graph, a trend characterized by strong ascending and descending steps, with a higher \ Delta in the quarter March, April and May, the maximum value of which occurs in April (3.04°C), while the lower it was found in December (-0.32°C).
\includegraphics[width=0.50\textwidth{}]{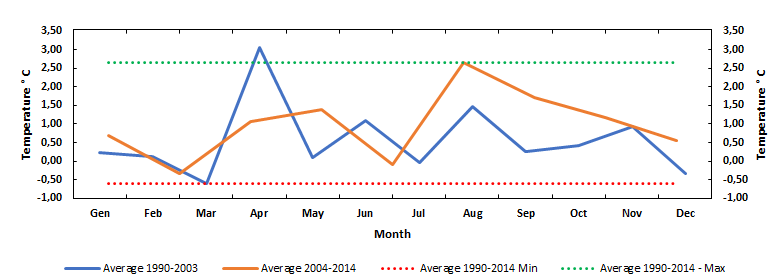} 
{\scriptsize \bf \textit{\textit{{\footnotesize Variation of average temperatures $\Delta$ in the municipal area of Montefiascone - VT}}}}
\subsection{Dew point} \normalsize In this session we describe the dew point obtained from an accurate analysis of the details referring to the municipal area. The average identified in the study period 1990-2014 shows at an annual level that in 14 years out of the 25 total studies, the values were above the average (equal to $8.91^\circ C$) as shown in figure 9 at the end of this article, while at a monthly level there are 6 months with values above the average, precisely in the period from May to October (see next image): \\ 
\includegraphics[width=0.49\textwidth{}]{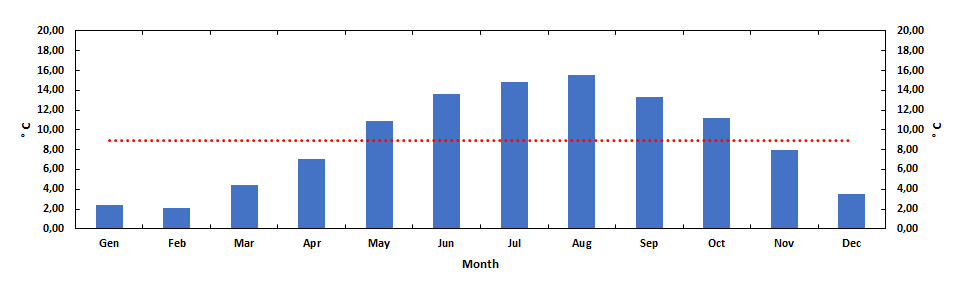} {\scriptsize \bf \textit{\textit{{\footnotesize Dew Point for months from 1990 to 2014}}}}. 
\subsection{Humidity} \normalsize As regards humidity, the study shows an annual value of $66.63\% $ calculated with the following formula::
\begin{equation}\frac{av}{am}\end{equation} 
{\footnotesize \textit{ {\bf av} = annual value, {\bf am} = average moisture 1990-2014}} \\ \\ The graph in Figure 8 shows the trend at the annual level while in the following graph it shows it at a monthly level. \\ \\
\includegraphics[width=0.49\textwidth{}]{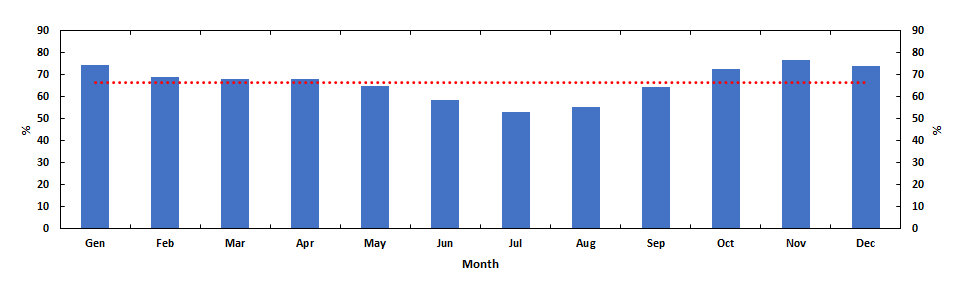} {\scriptsize \bf \textit{\textit{{\footnotesize Humidity Months from 1990 to 2014}}}}. \\ \\
The following graph, which highlights the monthly values, again referring to the 1990-2014 study period, highlights values above the average in the period between January and April, and after a below average summer quarter, return to rise above average again in the period October-December. \\ The annual study (figure 7 at the end of the document), highlights 13 years above the average, with the maximum value reached in 2002, and the remaining 12 years with values below the average, the lowest value of which is highlighted in 2012.
\subsection{Pressure} \normalsize The analysis of the data relating to pressure reports an average value in the period of the following study, equal to 1014.25 hPa. The following graph shows at a monthly level that January, June, July, August and October have values above the average, while the remaining months are positioned below the average, with March showing the lowest value.: \\ \\
\includegraphics[width=0.50\textwidth{}]{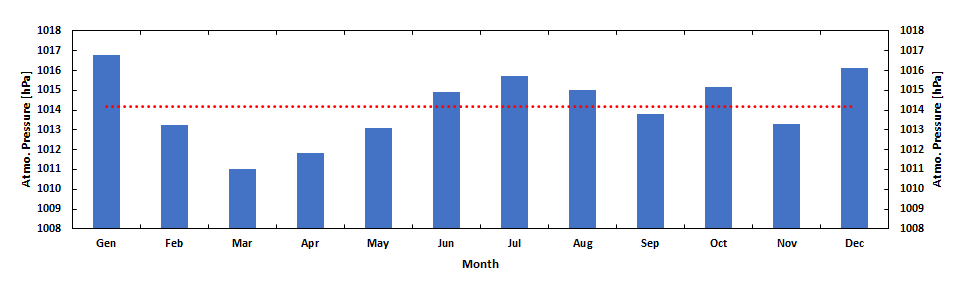} {\scriptsize \bf \textit{\textit{{\footnotesize Pressure Months from 1990 to 2014}}}}. \\ \\
Analyzing the data on an annual basis, as shown in Figure 8 at the end of the document, we note:
\begin{itemize}
\item an increase in atmospheric pressure from the annual average in the period: 1990-1992, 1998-2000, 2002 and 2005-2008;
\item a reduction in pressure from the annual average in the period: 1993-1997, 2001, 2003-2004, 2009-2014, with the lowest value reached in 2012.
\end{itemize}
\subsection{Day time and night time variation} \normalsize An accurate analysis of the daily $\Delta T$ variation of the temperature values between day and night is highlighted in this section and reported in Table 2.
The calculated value of $9,87^\circ C$ relating to the years 1990-2014 clearly evident in the following graph highlights 4 periods that are quite distinct from each other in the values of the temperatures found and precisely:
\begin{itemize}
\item an above average presence over the years: from 1990 to early 1994, from early 2000 to early 2008, with the greatest difference recorded in mid-2003, and in the period 2011 and mid-2012;
\item values below the average from the second half of 1994 until the end of 1999, from the beginning of 2008 and the end of 2010, with the lowest value recorded in mid-2009 and again from mid-2012 until the end of 2014.
\end{itemize}  
\includegraphics[width=0.50\textwidth{}]{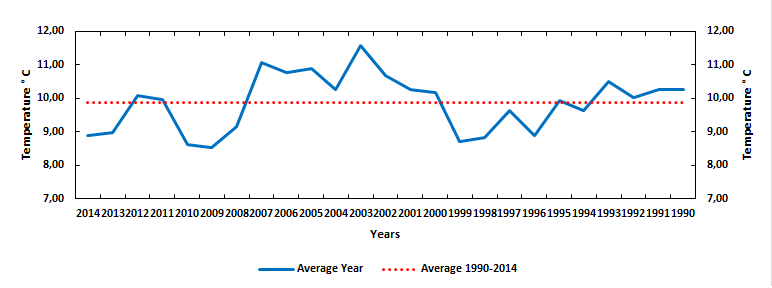}
{\scriptsize \bf \textit{\textit{{\footnotesize Day time and night time variations (solid line) and Average Day time and Night time variations 1990-2014 (dotted line)}}}}
\subsection{Rain} The analysis of the pluviometric data in the study period showed an average annual rainfall of 1503.45 mm in the municipal area, and an accumulation value for the period 1990-2014 equal to 37586.20 mm. Furthermore, an increase in intense and short-term rains emerged as clearly shown in figure 15 at the end of the document, which shows the rainfall trends for each month of the year referring to the values between 1990 and 2014. The following graph at a monthly level shows that all months have recorded an increase in the level of rainfall, with a maximum recorded in April and a minimum in July. 
\includegraphics[width=0.49\textwidth{}]{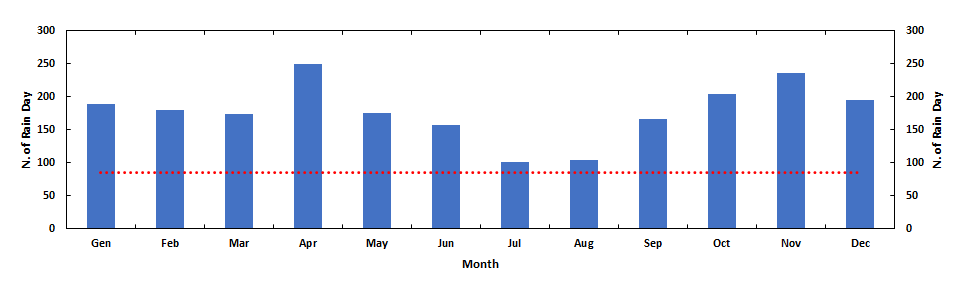} 
{\scriptsize \bf \textit{\textit{{\footnotesize Days Rain for Month from 1990 to 2014}}}}. \\ \\ These lead to an increase in hydro-geological risk as well as flash-food, rapid flooding of a morphologically limited area, due to the rapid “saturation” of the surface soil which is no longer able to absorb rain. Such episodes of intense precipitation can also determine phenomena of surface runoff of rainwater with a possible increase in flooding, but also risk of water pollution (pollutants from agricultural and road runoff). \\ The study period showed an abnormal increase in rainfall. The pie chart below shows the ratio between the total mm of rainfall measured in a year and the total rainfall measured in the period 1990-2014, according to this formula: 
\begin{equation}Total Rain=\frac{a}{b}x100
\end{equation} 
{\footnotesize \textit{{\bf a} = annual rainfall value, {\bf b} = total value of the rain period}} \\ \includegraphics[width=0.49\textwidth{}]{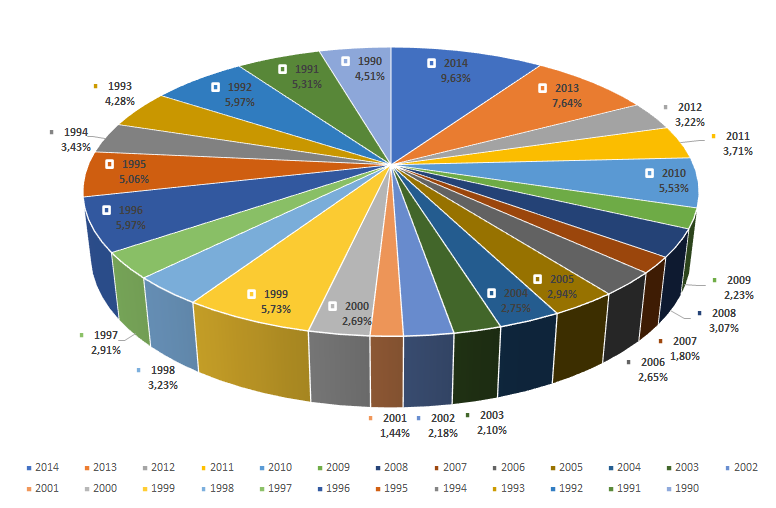}
{\scriptsize \bf \textit{\% of Rain for this study}}. \\
\\ In Figures 4 and 5 at the end of the document, we can see the total rainfall/year relating to this study, the values ​​of which are shown in table 3. The next pie chart highlights the type of phenomena found for this study. \\ \\
\includegraphics[width=0.5\textwidth{}]{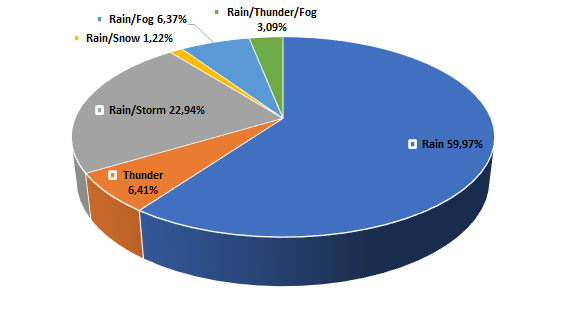}
{\scriptsize \bf \textit{Type of Rain for this study}}. \\ \\ The study highlighted annually:
\begin{itemize}
\item a considerable increase in rainfall in the years 2013 and 2014;
\item 10 years out of 25 study totals, in which rainfall was above average, with the highest value recorded in 2014;
\item 15 years in which rainfall was below average, with the lowest value recorded in 2001.
\end{itemize}
monthly:
\begin{itemize}
\item the month of April being the wettest.
\end{itemize}
in terms of number of rainy days per year:
\begin{itemize}
\item 15 years out of 25 in total in which the daily values exceeded the average of 85.44 mm, with the highest value reached in 2008;
\item 10 years out of 25 total in which the daily values were below the average, with the lowest value reached in 1999.
\end{itemize}
\subsection{Snow}
In the municipal area the presence of snow is not a rare event.
The nearby Vulsini mountains in fact determine a greater presence of snowy days and the municipal area under study clearly shows this being the municipality with the highest elevation on the southern shore of Lake Bolsena.\\ An essential element is obviously that there is the presence of the "wet flow" event that occurs when humid winds coming from the Tyrrhenian Sea flow over the coldest layer on the ground with winds coming from east-north-east. The Vulsini mountains are able to contain greater humidity in certain areas depending on the type of greater ventilation, favoring more rain and/or snow.\\
Obviously also the lake of Bolsena with its greater humidity present according to the winds, can generate the effect called "lake effect storm" which in particular cases becomes "lake effect snow".\\
The analysis of the data shows, as shown in the graph in figure 19, the number of snow days at an annual level, the average of the territory equal to 2 cm and the periods with the greatest snowfall that occurred in the years 1995-1996, 2001 and 2005 , with an average of 4 cm, while the lower values ​​occurred in the remaining years of the present study (1990, 1998-2000, 2006, 2011, 2014), interspersed with years in which no snow events occurred (1993-1994 , 1997, 2003-2004, 2007, 2009). The greatest snow accumulations, the average of which is equal to 32.72 cm and the total recorded is equal to 588.9 cm, were recorded in the years 1991-1992, 2006, 2010, 2011-2012 and 2013 with the highest values recorded nn 115 cm in the years 2012 and 2013. The minimum value recorded in 1990 equal to 5 cm, as shown in table 7 and in figure n. 20 at the end of the document.
\subsection{Fog} The analysis of the data showed that in the study period the average value of the days of the year with the presence of fog is equal to 36.92 as presented in the graph of figure 6, from which it can be verified that the values ​​have to decrease in the period between 1995 and 2011, finding in 2011 the year with the lowest number of days of fog in the municipal area (18 days), while the years between 1990 and 1994 recorded the highest values , with 1992 holding the highest value (78 days). Analyzing the data at a monthly level, we can verify as can be seen from the following graph, that the average of the days of fog is equal to 3.08, the months with the greatest number of fog days are October and November, while those with the lowest values ​are June and July. The analysis, as mentioned, showed a significant decrease in foggy days and associated with this, it is found that the fogs rarely manage to resist until the central hours of the day. In fact, the values ​​confirm the national trend in which the winter seasons are increasingly mild, with less cold nights which obviously lead to a deficit of cold and humidity, essential elements for the formation of fogs. \\ In the municipal area there is also, especially in winter and often at night, the phenomenon of "humidification fog", ie a dense mist due to the contrast between the colder air that blows on the surface of the lake and the temperature of the lake itself.
\includegraphics[width=0.49\textwidth{}]{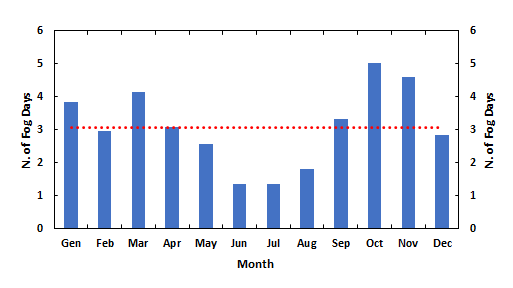}
{\scriptsize \bf \textit{Foggy month days from 1990 to 2014}}
\subsection{Wind speed} 
\normalsize The municipality of Montefiascone object of this study, being in the southern part of Lake Bolsena, the largest volcanic lake in Europe, is obviously affected by the cicroclimate that the lake itself creates.
\begin{itemize}
	\item At night the climate of Montefiascone is decidedly milder, being located on the southern part of Lake Bolsena, but often the mistral, the north wind and the north wind blow more intensely, reaching 80-100 km/h, albeit mitigating the climate and / or by stopping the thermal drop, the sensation of cold due to the increased ventilation is clearly evident.
	\item During the day the microclimate offered by the lake affects the wind, generating what in the area is called "Vattuta", a lake breeze coming from the north that keeps the southern shores of the lake and the territories therein cool, but this phenomenon occurs mainly in summer.
\end{itemize}
The ventilation of the territory can be viewed in the following figure generated by the Atlas of the Wind \cite{atlantevento:2013}, where the wind speed reached in the territory at 25 meters above sea level is highlighted. \\ In the municipal area, the study showed a constant decrease in wind speed over the years between 1990 and 2014.\\ Graph n.10 at the end of the document presents the minimum and maximum wind speed recorded in the municipal area. The data recorded a sharp decrease in the minimum values ​​in the period between January and October, with the first quarter well below the average of 6.74 km / h.
The maximum values ​​also recorded a below average trend, equal to 44.16 km / h in the month of January and in the period between March and October, while in November the most intense winds were recorded.\\
In table no. 4 you can view the daily wind values ​​reported in the following study, while in table 8 and in the monthly wind images relating to each single year of this study, they are visible in figure 21 (while on a daily level it is visible in figure 22) at the end of the document. \\ \\ 
\includegraphics[width=0.49\textwidth{}]{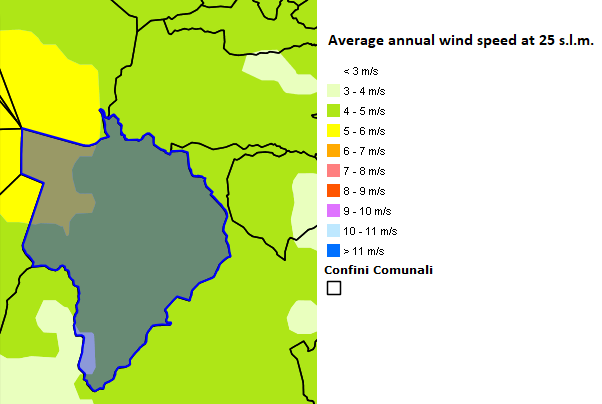} \\ 
{\scriptsize \bf \textit{Wind Speed Atlant of Montefiascone - VT \cite{atlantevento:2013}}}
\section{Geological and Earthquake data} \normalsize In this section we describe the Geological and Earthquake data. \\
% Inser Geological Map\%
\includegraphics[width=0.49\textwidth{}]{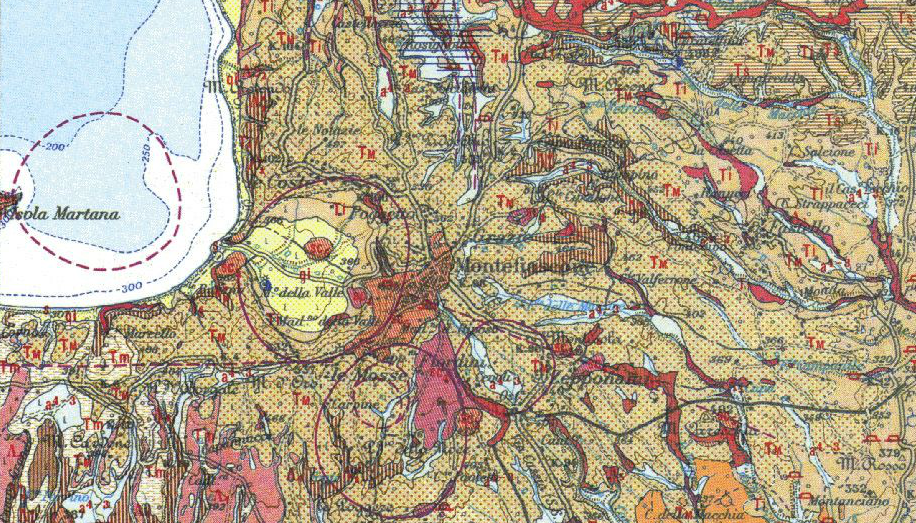} \\
{\scriptsize \bf \textit{Distribution of the main neogenic and quaternary basins of the Northern Apennines. 1=main overlapping fronts; 2=main faults at the edges of the basins; 3=transverse tectonic lines; 4=minor faults at the edge of the basins (BOSSIO et all. 1992).}} \cite{mappageologica:2013}
\subsection{Morphological Lines}
\normalsize Morphological characteristics and structure of the survey area were strongly influenced both by the nature of the outcropping rocks and by the exogenous and endogenous processes that have followed and alternated in the last millions of years.\\
Predominant are the landscapes resulting from the spreading outcrop of volcanic rocks belonging mainly to the Vulsino Volcanic District, which have in fact conditioned the topography of the area, characterizing it with a series of hilly reliefs (maximum altitudes around 600-700m above sea level) that are more emission centers, and which alternate with large volcano-tectonic depressions, the largest of which is occupied by Lake Bolsena. The positive forms are represented by numerous scoria and ash cones (for example Montefiascone and Valentano), while the most evident negative forms are the large elliptical or sub-circular calderas of Latera and Montefiascone.\\
Rather steep slopes, in correspondence with the most recent volcano-tectonic structures (edges of calderas, faults and fractures) and of the outcrop of lithoid rocks (lava flows), alternate with softer slopes, in correspondence with the lithotypes less resistant to erosion (less coherent pyroclastic products) and large structural surfaces (plateauxignimbritici).\\
The action of running waters and the processes connected with the eustaticowurmian uplift, have affected generally narrow and deep valleys, subsequently remodeled and partially covered by alluvial deposits.
\subsection{Geological Characteristics}
\normalsize As already mentioned, volcanic and pyroclastic rocks are the most widespread ones in the area under examination within which even those belonging to the sedimentary substrate emerge, albeit marginally.\\ The following description is therefore proportionate to the outcrop extension and the importance that both play for the purposes of this study.
\subsection{Volcanic Formations}
\normalsize The geological-deformation history of the study area of this article is closely connected with that of the Lazio volcanic region which must be framed within the context of the tectonic-dynamic evolution of central Italy. \\ The geological evolution of this area is the result of the geodynamic processes and the extension tectonics of the belt between the Apennine chain and the Tyrrhenian coast, following the Apennine orogeny and active in the period between the upper Pliocene up to just under 50,000 Years ago. \\ Starting from the Pliocene, an intense magmatic activity begins that involves, along the Apennine route, a vast range from Monte Amiata to the Campania region.\\ The volcanoes of Lazio, present in the area of the present study, belong to two clearly distinct magmatic series: the first includes the acid, rhyolithic and riodacitic volcanism of the Cimini Mountains, the Tolfa Mountains and the Ceriti Mountains and has an older age ( between about 2 and 1 Ma); the second series includes the Vulsino, Vicano, Sabatino and Colli Albani groups, and shows a distinctly alkaline-potassium character and began its active state 0.8 Ma ago, continuing up to the present time. The following figure identifies this division.
\includegraphics[width=0.49\textwidth{}]{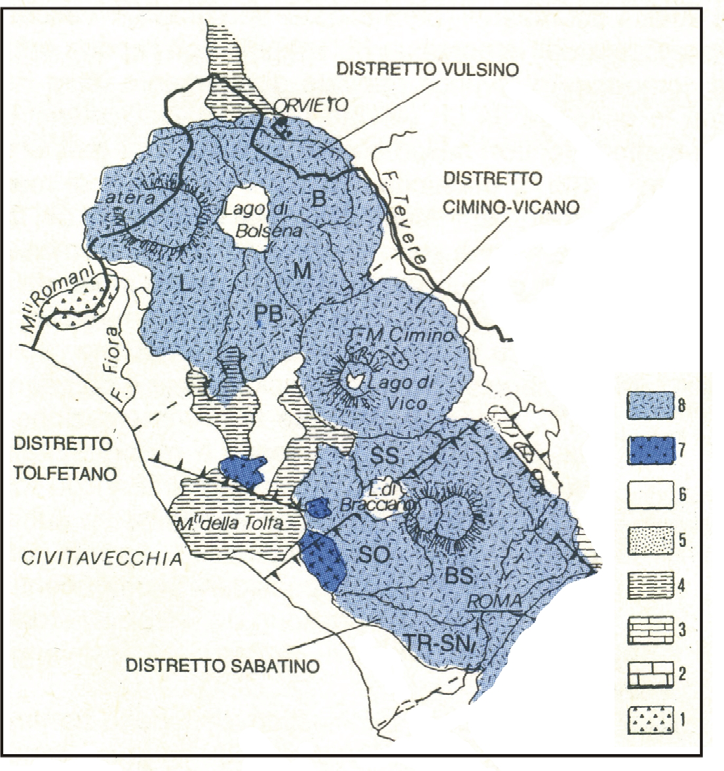} \\
{\scriptsize \bf \textit{Main volcanic complexes of northern Lazio. Legend: 1: rocks of the metamorphosed base; 2: sediments from the Lazio-Abruzzo platform; 3: sediments of the Umbrian Marche basin; 4: allochthonous sediments of the Ligurian and subligur complex; 5: flyschoid allochthonous sediments; 6: neo-autochthonous sandy-clayey-gravelly sediments; 7: volcanic districts with acid to intermediate chemism; 8: volcanic districts with a potassic to highly potassic character (PB: Paleobolsena volcanic complex; B: Bolsena volcanic complex; M: Montefiascone volcanic complex; L: Latera volcanic complex; SW: western sector activity; SS: activity of the northern sector; BS: Sacrofano-Baccano complex; TR-SN: pyroclastic flow of red tuff with black slag) - (modified from Italian Geological Society, 1993)}}.
\subsubsection{Vulsino Volcanic District} 
\normalsize The district to which this study belongs falls within the Vulsini District whose activity, in accordance with Beccaluva et alii (1991) \cite{Beccakuva:1991}, can be linked to the partial fusion and heterogeneous enrichment of a source that can be localized in the mantle. The Vulsino Volcanic District is the northernmost of the volcanic districts of Lazio and occupies an area of about $2200 km^{2}$ extending between the Fiora River and the Tevere River. The area sees the prevalence of sub-area activity mainly of an explosive nature, and is located at the intersection of a complex system of faults in the Apennine and anti-Apennine directions. This apparatus was the protagonist of the most impressive and extensive volcanic events in the region. \\ During its activity there was the emission of a great variety of volcanic products (ignimbrites, lavas, pyroclasticites of various kinds) attesting an activity that began in the Pleistocene, lasted until very recent times and is still in progress in the form hydrothermal and sulfataric. \\ The sedimentary substrate is made up of neo-autochthonous post-orogenetic deposits of the upper Miocene and the Pleistocene, which cover the flyschoid units and the Mesozoic carbonate sequences of the Tuscan and Umbrian-Marche domains \cite{Nappi G.:ref1995}. Within the evolution of the Vulsino Volcanic District, five areas or volcanic complexes have been distinguished: Paleobolsena, Bolsena, Montefiascone, Latera and Neobolsena \cite{Nappi G.:ref1995} \cite{Nappi G.:ref1998} \cite{Nappi G.:ref2004}), with multiple mechanisms and eruptive scenarios: the spectrum of explosive activities, which includes those of the Hawaiian, Strombolian, Plinian, hydromagmatic and surtseyan types is in fact almost complete. \\ The deposits related to these eruptive mechanisms are represented by welded slag, slag cones, layers of pumice, ignimbrites, surges, lapilliaccretionals, etc. Also the products of the effusive activity are well represented and reflect a broad compositional spectrum, ranging from the series leucitic to shoshonitic. The most differentiated products are present in the areas of Paleobolsena and Bolsena, while the area of ​​Montefiascone, in correspondence with which the magma chamber is located in the upper part of the carbonate base, is characterized by less differentiated products. This study examines the area of Montefiascone.
\includegraphics[width=0.49\textwidth{}]{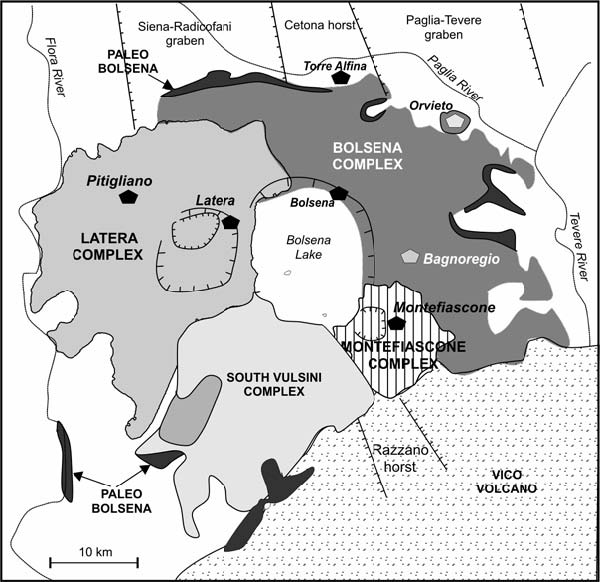} \\
{\scriptsize \bf \textit{Schematic geological map of the Vulsini area - source:https://www.alexstrekeisen.it/provincie/vulsini.php}}
\subsection{Volcanic area of Montefiascone}
\normalsize The initial activity of the Montefiascone complex is mainly identified with the Ospedaletto eruption \cite{Nappi G.:ref1995}, characterized by a Plinian eruptive column about 20 km high \cite{Nappi G.:ref1994a} which gave rise to deposits of pumice covering a very large area in the eastern and southern sectors.\\ The eruptive scenarios of the Montefiascone volcanic area were influenced by the structure of the carbonate base. NNO-SSE fractures have in fact determined an intense pre and post-calderic effusive activity in the central band; in the southern one, it was an O-E fault that represented the supply route for scoria cones and vast lava flows.\\ The explosive activity in the Montefiascone area was very intense and mainly of the hydromagmatic type.\\
The Montefiascone Ignimbrite and the formation of the relative caldera are connected to the eruption richer in energy\cite{Nappi G.:ref1986}. The basal ignimbrite, which crops out in the southern and eastern sector, appears as a massive deposit from light gray to dark gray.\\
A second central explosive eruption determines the formation of Lava Drop Ignimbrite. A further collapse of a part of the northern sector of the caldera itself, while a resumption of activity with hydromagmatic phases, gave rise to the chaotic collapse of the eastern sector\cite{Nappi G.:ref1986}.\\
Among the lava products of the Montefiascone complex there is a prevalence of leucitites, tephra and basanites.
\subsection{The Sedimentary Units}
\includegraphics[width=0.49\textwidth{}]{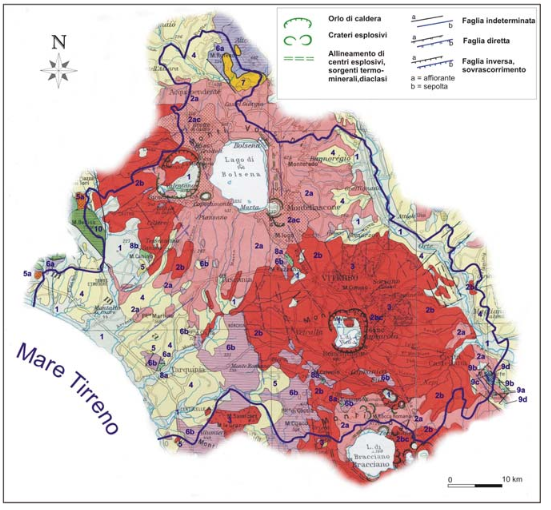} 
{\scriptsize \bf \textit{Geological diagram of the Province of Viterbo. Legend: 1: undifferentiated continental and marine deposits, Quaternary. Volcanic units-2a: undersaturated volcanites; 2ac: in hydromagmatic facies, middle-upper Pleistocene; 2b: intermediate vulcanites; 2bc: in hydromagmatic facies, middle-upper Pleistocene; 3: acid volcanites, lower Pleistocene. Postorogenic unit-4: marine terrigenous units, Pliocene – Lower Pleistocene; 5: Unit of Poggio Terzolo, Messinian; 5a: “Bismantova Group”, Lower Miocene – Middle Miocene. Internal Ligurian and Australoalpine Units-6a: S. Fiora Formation, Upper Cretaceous – Eocene; 6b: Flysch of the Tolfa, Late Cretaceous – Eocene; 7: Formation of clays with Palombine limestones and ophiolites, Lower Cretaceous. Tuscan Units - 8a: Macigno Auct., Upper Oligocene - Miocene; 8b: flint limestone and marl with radiolarians, Liassic - Miocene. Non-metamorphic Tuscan Roof Units - 9: massive limestone, red ammonite, flint limestone, jasper, polychrome schist, Jurassic - Eocene. Umbrian-Marche units-9a: bisciaro, scaglia, marl with fucoids Auct., Lower Cretaceous - Lower Miocene; 9b: flint and marl limestones, Lower Cretaceous – Middle Liassic; 9c: pelagic limestones, middle-upper Jurassic; 9d: platform limestones and dolomites, Triassic. Unit of the metamorphic base - 10: Unit of Boccheggiano, Permiano (from the National Research Council, 1987; modified by Chiocchini and Madonna, 2004).}} \\ \\
\normalsize Sedimentary rocks emerge in isolated strips, interspersed with volcanites, generally in correspondence with tectonic culminations and fluvial incisions and more extensively, in the south-eastern sector of the area studied, on the edge of volcanic rocks.
The geological reconstruction of the units referable to the sedimentary substrate was carried out with reference to the available bibliography.
They are essentially attributable to infacies of flysch, allochthonous deposits (Upper Cretaceous-Oligocene) and marine deposits of the post-orogenic cycle (Pliocene-Pleistocene).\\
The flyschoid units are made up of marl, clay, marly limestone and sandstone from the Late Cretaceous to Oligocene.\\
In the study area they are represented by the "Flysch della Tolfa" sensu Chiocchini e Madonna, 2003\cite{Chiocchini:2003} which according to Buonasorte et alii (1988)\cite{Buonasorte:1987} belongs to the internal Austroalpine domain, while Abbate e Sagri (1970)\cite{Abbate:1970} place it in the Ligurian domain.\\
The Tolfa Flysch has been framed as a Series or Comprehensive Succession, aged between the Late Cretaceous and the Middle Eocene, by Alberti et alii (1970)\cite{Alberti:1970} and by Bertini et alii (1971).\\
According to Civitelli e Corda (1993)\cite{Civitelli:1993} the succession consists of three members: one clayey-calcareous, one calcareous-marly (with interspersed varicolored argillites, pietraforte and red marl) and one arenaceous.\\
Post-orogenic units comprise three groups of sequences:
\begin{itemize}
\item the first includes units aged between the Messinian and the lower Pliocene base;
\item the second includes units of the Pliocene;
\item the third includes Quaternary units, widely represented on the margins of the Lazio volcanic blanket and inside it, in correspondence with valley engravings.
\end{itemize}
Their deposition is connected to a predominantly marine sedimentary cycle that affected the Tyrrhenian side of the central-northern Apennines from the Messinian to the Quaternary.\\
The Pliocene units emerging in the study area, sometimes rich in fossil fauna, belong to the Units of the Tarquinia basin (Chiocchini e Madonna, 2003\cite{Chiocchini:2003}). Alberti et alii (1970)\cite{Alberti:1970} recognize a succession composed at the base of pelites and conglomerates of the lower-middle Pliocene age, on which sands and conglomerates of the middle-upper Pliocene rest. These units are part of the mainly marine neo-autochthonous cycle and are found in strips south of Tuscania and in the Monte Romano area, overlapping the Flysch della Tolfa. Quaternary units are also represented in the study area, consisting of continental travertine deposits (from the Pleistocene-Holocene age), which emerge in the areas of Bullicame and Bagnaccio (west of Viterbo) and in LaRocca, and from Holocene alluvial and lake deposits. , outcropping in the valley engravings of the rivers and along the edges of Lake Bolsena.
\subsection{Geochemistry of Vulsian rocks} 
\normalsize The chemical compositions of vulsine rocks are characterized by a high concentration of K, Rb, Th, U and other rare elements. An easy way to show the chemical composition of rocks is by using so-called spider diagrams, in which the concentrations of the individual elements of the rocks are divided by the concentrations of the same elements of a reference material, such as the average composition of the Earth or that of chondritic meteorites.\\
This method provides different curves for the various rocks that show in a very effective visual way the difference between the composition of the rocks themselves compared to the average of the Earth, for the various chemical elements.\\
When using this method for vulsine rocks, it is noted that the concentrations of radioactive elements (K, Rb, Th, U) are three orders of magnitude higher than the Earth's averages. The comparison with other terrestrial rocks such as oceanic basalts (the most abundant igneous rocks on Earth) and limestones, better highlight the geochemical anomalies of Vulsian volcanism.\\
This means that the natural radioactivity in the Vulsini is much higher than in the regions of the planet where there are normal basalts or even other types of rock such as limestone or sandstone. Other chemical elements, such as fluorine (F) and arsenic (As), are also highly enriched in Vulsian rocks. This has important effects on the water quality of this region, as the high concentration in the rocks of these potentially toxic elements is transmitted to the waters.\\
For example, it should be remembered that the upper limit of the concentration of F in drinking water is only 3 parts per million (3 milligrams / liter).\\
One of the main problems of Vulsin magmatism is to understand the causes of the strong enrichment in K, Rb, Th, U, and other rare elements. According to the most recent hypotheses, this is the effect of two processes:
\begin{itemize}
\item one that occurs inside the earth's mantle where magmas are generated;
\item the other concerns the magmas during their ascent to the surface, when they remain for more or less long times inside the magma chambers below the volcano.
\end{itemize}
\normalsize According to the theory of plate tectonics, the earth's crust and part of the upper mantle (the lithosphere) form rigid blocks that move reciprocally with a relative speed of a few cm per year. Most of the magmas erupted on the planet are generated in the upper mantle. Magmatism is concentrated either in areas of diverging plates (e.g. mid-ocean ridges) where the plates move away from each other, or in areas of convergent plates where the plates approach and underflow with respect to each other (lithospheric subduction ). A small number of volcanoes (for example those of Hawaii) are located in areas inside the lithospheric plates, in correspondence with thermally and / or compositionally anomalous points, called "hot spots" (Figure 15). The upper mantle is composed almost entirely of solid materials. However, in areas of lithospheric divergence or in areas of convergence above the subduction zones, the upper mantle can locally melt, creating large quantities of magmas. The composition of magmas strongly depends on that of the rocky mantle from which they were generated. The upper mantle generally contains low amounts of K, Rb, Th, U, etc.\\ However, these elements are present in higher anomalous concentrations in the upper mantle areas above the subduction areas. Consequently, the magmas produced in these areas have a higher concentration of K, Rb, Th, U, etc. than those generated along the boundaries of the diverging plates.\\ The enrichment of chemical elements in the mantle above the subduction areas is due to the introduction through the subducting plate of sediments or other materials of the crust that contain high concentrations of K, Rb, Th, U, etc.\\ It is generally accepted that the magma erupted in the Vulsini and other volcanoes of central and southern Italy formed in the upper mantle above a subduction zone. In this case, the subducting plate is made up of the Adriatic lithosphere which converges towards the Italian peninsula and plunges under the Apennine chain.\\ The crust material transported by the Adriatic plate produces an enrichment of the upper mantle in K, Rb, Th, U and in other elements such as F and Cl (chlorine) and CO2 (carbon dioxide). The fusion of this chemically anomalous mantle gave rise to the magmas of the Vulsini and other volcanoes of central Italy.\\ A second process of geochemical enrichment of magmas occurs when they are stationed in the magma chambers present under many Italian volcanoes at a depth of a few kilometers.\\
In these reservoirs the magmas derived from the mantle lose heat and partially crystallize, the residual liquids become very rich in K, Rb, Th, U, etc. \\ \\
\includegraphics[width=0.49\textwidth{}]{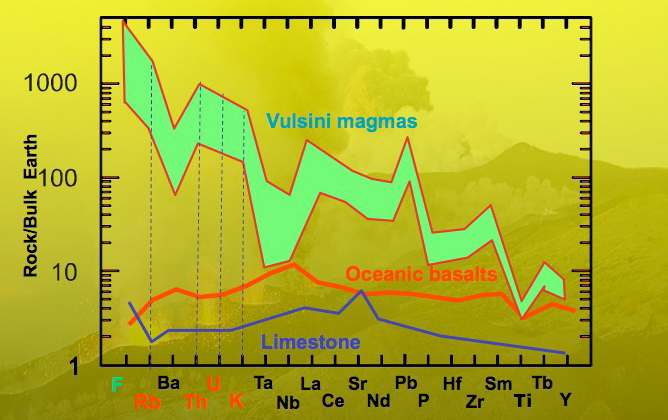} \\
{\scriptsize \bf \textit{Chemical composition diagram of rocks in the study area. Image from: http://www.orviamm.com/medias/files/peccerillo-geologia-e-origine-dei-vulcani-vulsini.pdf \cite{Pecceriello A:ref2}}}
\subsection{Earthquake and Seismic Response Spectrum}
\normalsize The municiplay of Montefiascone is characterized by a territory with medium-hight seismicity, \\ \includegraphics[width=0.49\textwidth{}]{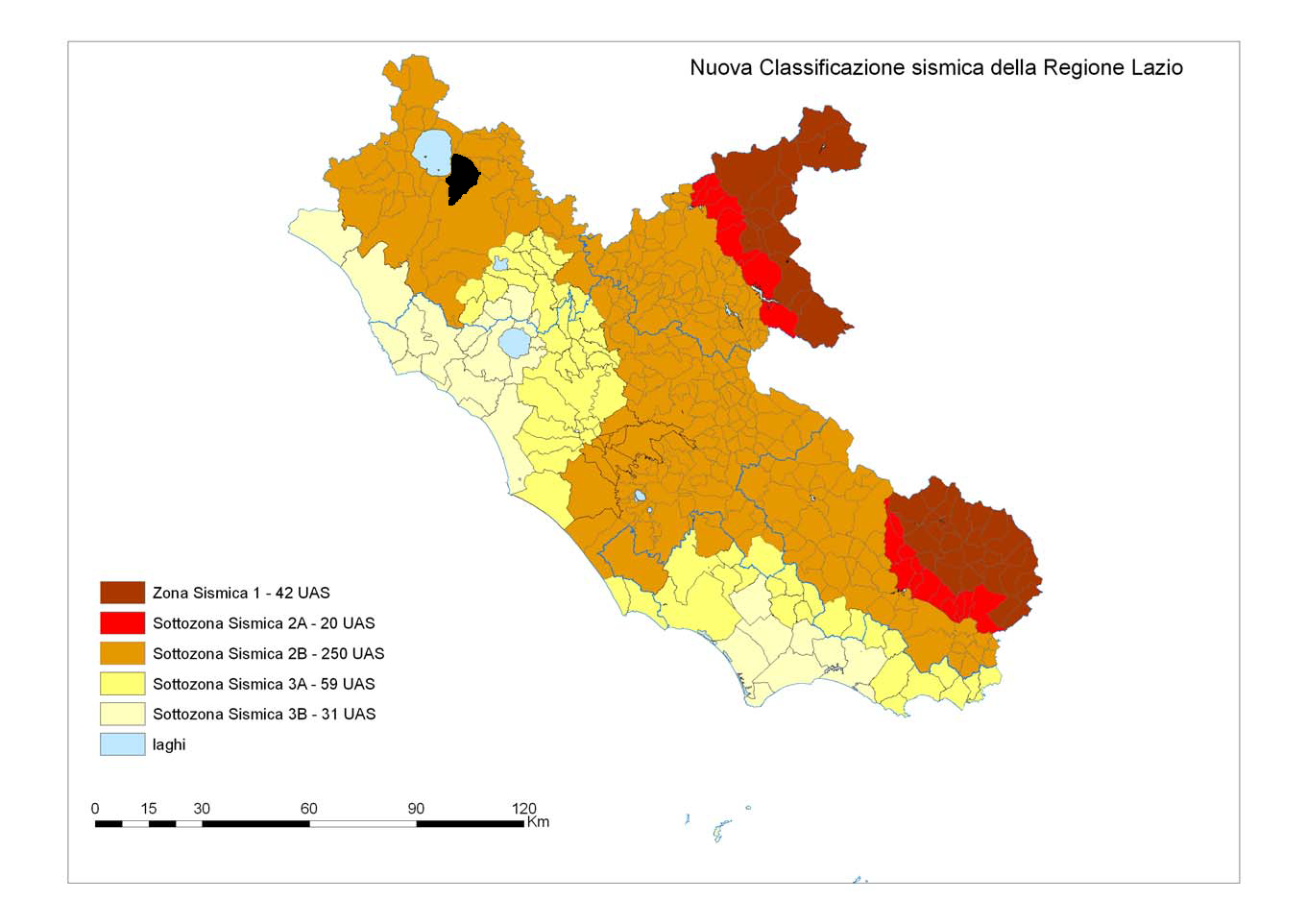} \\
{\scriptsize \bf \textit{Seismic Hazard Map (MPS) of Lazio. The acceleration values refer to a return of 475 years (INGV2004). The city of Montefiascone is highlighted in black on the map.}} \cite{INGV:2004} \\ \\
in fact in table A attached to the Decree of 01/14/2008 draw up by the Ministry of Infrastructures, the seismic hazard  estimates are highlighted and from these the response spectrum was determined elastic (horizontal and vertical) of the seismic actions highlighted in the following graph.
\includegraphics[width=0.49\textwidth{}]{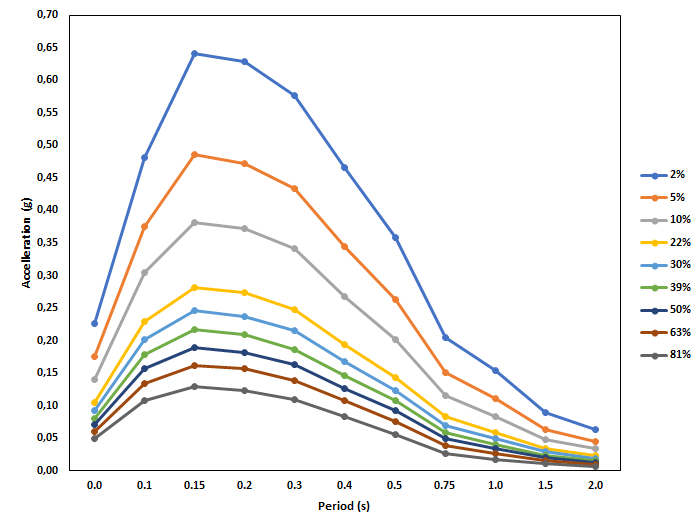} \\
{\scriptsize \bf \textit{The spectra indicate the calculated shaking values for 11 spectral periods, ranging from 0 to 2 seconds. The PGA corresponds to the period of 0 seconds. The graph relates to the median estimates (50th percentile) proposed by hazard model. The different spectra in the graph relate to different probabilities of overflow (PoE) over 50 years.}} \cite{INGV:2004:ref2} \cite{Stucchi:ref1} \\ \\
In next figure we can see the number of Earthquake in Montefiascone, and table evidence the number of Earthquake event by year. \\ \\
\includegraphics[width=0.5\textwidth{}]{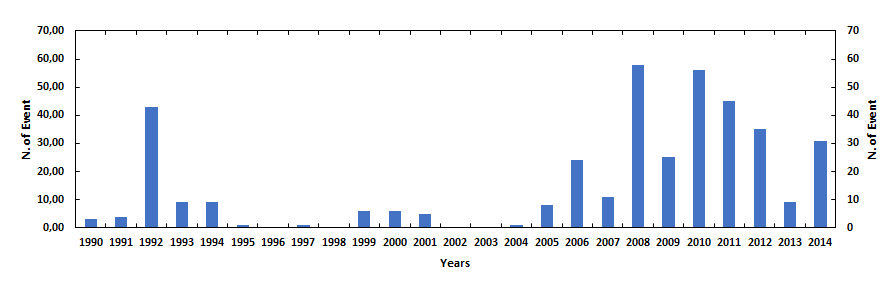}
{\scriptsize \bf \textit{Number of Earthquake for 1990 to 2014}} \\ \\
% Table of Seismic Event
% Table generated by Excel2LaTeX from sheet 'Tabella Riassuntiva '
\begin{tabular}{|r|c|c|c|}
\hline
\scriptsize \textbf{Year} &\scriptsize \textbf{N. of Events} &\scriptsize \textbf{ Year} & \scriptsize \textbf{N. of Events}\\
\hline
\scriptsize 2014 & \scriptsize 6 & \scriptsize 2013 & \scriptsize 2 \cr
\scriptsize 2012 & \scriptsize 1 & \scriptsize 2011 & \scriptsize 4 \cr
\scriptsize 2010 & \scriptsize 0 & \scriptsize 2009 & \scriptsize 1 \cr
\scriptsize 2008 & \scriptsize 0 & \scriptsize 2007 &\scriptsize 0 \cr
\scriptsize 2006 & \scriptsize 2 &\scriptsize 2005 & \scriptsize 1 \cr
\scriptsize 2004 & \scriptsize 0 & \scriptsize 2003 & \scriptsize 0 \cr
\scriptsize 2002 & \scriptsize 1 & \scriptsize 2001 & \scriptsize 0 \cr
\scriptsize 2000 &\scriptsize 3 & \scriptsize 1999 & \scriptsize 0 \cr
\scriptsize 1998 & \scriptsize 0 & \scriptsize 1997 & \scriptsize 0 \cr
\scriptsize 1996 & \scriptsize 0 &\scriptsize 1995 & \scriptsize 0 \cr
\scriptsize 1994 & \scriptsize 1 & \scriptsize 1993 & \scriptsize 2 \cr
\scriptsize 1992 & \scriptsize 1 &\scriptsize 1991 & \scriptsize 0 \cr 
\scriptsize 1990 &\scriptsize 0 & & \\
\hline
\end{tabular} \\ \\
\scriptsize {\bf {Number of Earthquake in Montefiascone by Year}} \\ \\
\normalsize The study evidence that the major number of Earthquake are production on deep from 5 to 10 Km. Next graphics evidence this. \\ \\
\includegraphics[width=0.5\textwidth{}]{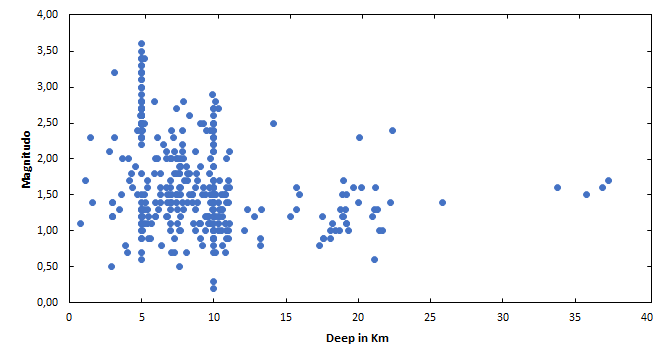} \\
{\scriptsize \bf \textit{Magnitudo/Deep for 1990 to 2014}}
\section{Conclusion} 
\normalsize We presented for the first time an analysis of geological and longterm temperature data directly obtained from Montefiascone local meteorological site, inside urban concentration and well above the inversion layer. \\
The data analysis above all those related to annual rainfall, to their regimen and the change of these indices over time indicates that the area is recorded increase the contribution of rainfall and its concentration in the most rainy periods. \\
The extremes of this trend over time if confirmed, would lead to a radical change in climate of the area that would no longer be characterized by the absence of the dry season, but rather by a rainy season with heavy and often heavy rainfall and a dry season with characteristics similar to those of the dry seasons of the Mediterranean climate. \\ The inversion of Montefiascone arouses contrasting effects among citizens, as it is a phenomenon particularly popular in the middle of a hot summer when at least in the early morning you are able to get much lower temperatures and refreshing, while in winter it is seen as a phenomenon inconvenient dates the low temperatures and possible fog and frost that usually occur in parallel. \\
Given the geology structure of the rocks and the territory, and highlighted the amount of railfallin the period presented in this study (highlighted in section 3.7 and in Table 3), it is recommended further attention in the monitorng of climatic phenomena, in order to prepare the necessary and appropriate measure to protect the population and existing infrastructure.
\section{Acknowledgments}
\normalsize We would like to thank Dr.ssa Filomena Barra for her helpful suggestions and support to made the paper more complete. The constructive comments are highly appreciated. 
\end{multicols}
\newpage
% Inser Area of Montefiascone\%
\begin{figure}
\begin{center}
\includegraphics[width=0.8\textwidth]{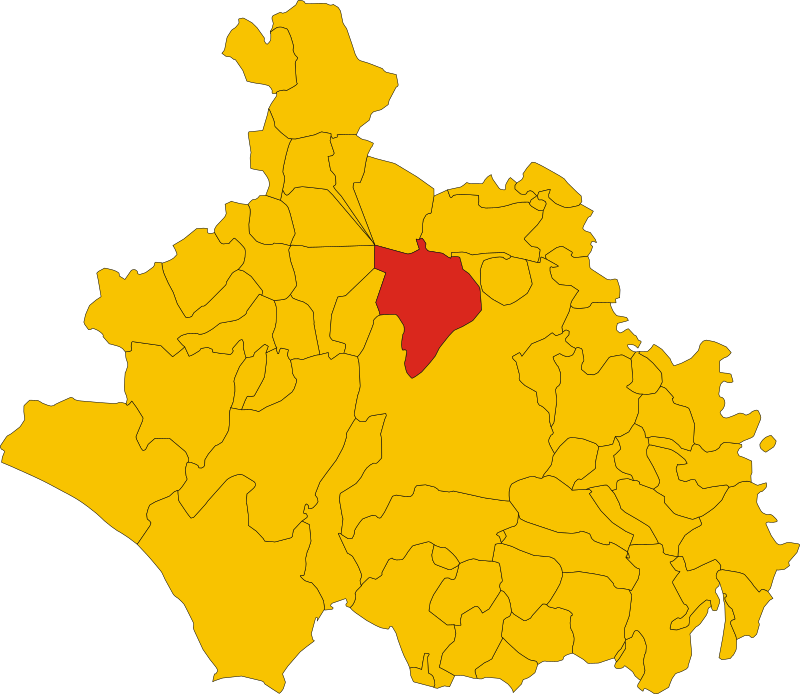}
\caption{Identification of Montefiascone in Viterbo's  province}
\end{center}
\end{figure}
\clearpage
\begin{table}
\normalsize
\caption{Comparison of temperature on decadal scale from Jannary to December} 
\tiny  \rotatebox{90}{
\begin{tabular}{|c|c|c|c|c|c|c|c|c|c|c|c|c|c|c|c|c|c|c|c|c|c|c|c|c|c|}
	\hline
	\multicolumn{ 2}{|c}{Year}& \multicolumn{ 2}{|c}{Jan} & \multicolumn{ 2}{|c}{Feb} & \multicolumn{ 2}{|c}{Mar} & \multicolumn{ 2}{|c}{Apr} & \multicolumn{ 2}{|c}{May} & \multicolumn{ 2}{|c}{Jun}& \multicolumn{ 2}{|c}{Jul} & \multicolumn{ 2}{|c}{Aug}& \multicolumn{ 2}{|c}{Sep} & \multicolumn{ 2}{|c}{Oct}& \multicolumn{ 2}{|c}{Nov} & \multicolumn{ 2}{|c|}{Dec}\\
	\hline
	\multicolumn{ 1}{|c}{} && Min &        Max &        Min &        Max &        Min &        Max &        Min &        Max &        Min &        Max &        Min &        Max &        Min &        Max &        Min &        Max &        Min &        Max &        Min &        Max &        Min &        Max &        Min &        Max\\
	\hline
	\bf{2014} && -2,00 & 16,00 &   1,00 & 20,00 &    -  & 21,00 &  4,00 & 22,00 &  6,00 & 26,00 & 10,00 & 34,00 & 14,00 & 33,00 & 14,00 & 31,00 & 12,00 & 28,00 &  6,00 & 27,00 &  2,00 & 20,00 &  -2,00 & 19,00 \\
	\bf{2013} && -2,00 & 16,00 &  -3,00 & 15,00 & -3,00 & 19,00 &  4,00 & 26,00 &  6,00 & 24,00 & 11,00 & 34,00 & 14,00 & 35,00 & 16,00 & 38,00 & 10,00 & 32,00 & 10,00 & 24,00 & -1,00 & 22,00 &  -2,00 & 14,00 \\
	\bf{2012} && -5,00 & 15,00 &  -7,00 & 18,00 &  2,00 & 24,00 &  1,00 & 26,00 &  6,00 & 28,00 & 12,00 & 35,00 & 17,00 & 37,00 & 16,00 & 38,00 & 10,00 & 32,00 &  1,00 & 27,00 &  3,00 & 19,00 &  -5,00 & 14,00 \\
	\bf{2011} && -4,00 & 16,00 &  -1,00 & 19,00 & -4,00 & 19,00 &  2,00 & 25,00 &  7,00 & 30,00 & 14,00 & 33,00 & 12,00 & 35,00 & 14,00 & 36,00 &   -   & 33,00 &  5,00 & 28,00 &  1,00 & 20,00 &  -1,00 & 19,00 \\
	\bf{2010} && -3,00 & 13,00 &  -3,00 & 16,00 & -2,00 & 18,00 &  3,00 & 23,00 &  6,00 & 25,00 & 10,00 & 30,00 &    -  & 35,00 & 13,00 & 34,00 &  9,00 & 28,00 &  4,00 & 25,00 & -1,00 & 23,00 &  -7,00 & 17,00 \\
	\bf{2009} && -3,00 & 13,00 &  -4,00 & 15,00 &    -  & 19,00 &  6,00 & 23,00 & 10,00 & 32,00 & 13,00 & 32,00 & 16,00 & 35,00 & 17,00 & 35,00 & 12,00 & 31,00 &  1,00 & 26,00 &  3,00 & 21,00 &  -5,00 & 17,00 \\
	\bf{2008} && -2,00 & 19,10 &  -6,00 &   -   & -0,80 & 19,50 &  1,20 & 21,10 &  6,40 & 29,50 & 12,00 & 32,00 & 16,00 & 34,00 & 16,00 & 35,00 &  7,00 & 35,00 &  4,00 & 26,00 &  1,00 & 22,00 &  -1,00 & 14,00 \\
	\bf{2007} && -2,00 & 18,50 &  -1,40 & 21,00 &  0,80 & 21,20 &  5,00 & 26,70 &  8,20 & 31,90 & 12,00 & 35,10 & 11,80 & 36,90 & 13,60 & 36,90 &  9,00 & 30,10 &  4,20 & 30,00 & -0,80 & 19,70 &  -3,40 & 15,30 \\
	\bf{2006} && -5,30 & 15,50 &  -4,40 & 18,10 & -2,40 & 20,00 &  3,00 & 23,70 &  6,20 & 29,30 &  7,60 & 34,70 & 13,60 & 38,00 & 13,80 & 33,70 & 11,60 & 32,70 &  5,80 & 28,50 & -1,00 & 20,30 &   1,00 & 19,50 \\
	\bf{2005} && -3,80 & 14,90 &  -4,00 & 15,70 & -6,60 & 20,70 &  2,40 & 26,50 &  7,40 & 31,30 &  5,20 & 36,90 & 12,40 & 39,00 & 10,80 & 33,70 & 11,00 & 32,70 &  6,80 & 23,10 &  3,50 & 20,00 &  -4,80 & 14,10 \\
	\bf{2004} && -5,40 & 16,10 &  -2,00 & 17,50 & -2,30 & 20,70 &  2,80 & 23,00 &  3,80 & 25,00 &  9,40 & 33,10 & 11,40 & 37,70 & 12,80 & 34,70 &  9,40 & 33,70 &  8,80 & 26,50 & -2,00 & 25,30 &  -2,40 & 18,90 \\
	\bf{2003} && -4,00 & 17,90 &  -4,20 & 16,70 & -1,40 & 20,10 & -5,40 & 27,90 &  8,40 & 32,30 & 14,40 & 36,70 & 15,00 & 38,10 & 16,00 & 39,10 & 10,40 & 31,00 &  3,60 & 26,90 &  0,60 & 19,10 &  -3,80 & 17,00 \\
	\bf{2002} && -7,00 & 15,00 &  -1,00 & 16,30 & -0,80 & 25,30 &  1,00 & 26,80 &  6,40 & 27,30 & 10,60 & 37,10 & 11,80 & 34,50 & 10,40 & 32,50 &  7,80 & 26,50 &  6,20 & 23,00 &  0,80 & 21,70 &  -0,20 & 14,90 \\
	\bf{2001} &&  1,20 & 15,70 &  -4,80 & 16,30 &  1,00 & 22,10 & -2,00 & 24,30 &  9,00 & 28,50 &  8,00 & 34,30 & 13,60 & 34,50 & 15,20 & 35,70 &  7,80 & 28,50 &  5,40 & 26,70 & -1,40 & 21,10 &  -4,20 & 15,30 \\
	\bf{2000} && -5,00 & 13,00 &   1,00 & 16,00 & -1,00 & 18,00 &  3,00 & 25,00 & 10,60 & 27,00 & 11,60 & 31,70 & 10,00 & 36,10 & 12,80 & 38,10 &  9,00 & 29,90 &  8,00 & 27,90 &  3,00 & 20,30 &  -1,00 & 17,30 \\
	\bf{1999} && -5,00 & 16,00 &  -4,00 & 15,00 &  1,00 & 21,00 &  2,00 & 21,00 & 10,00 & 29,00 & 15,00 & 29,00 & 14,00 & 32,00 & 16,00 & 36,00 & 14,00 & 29,00 &  6,00 & 25,00 &  1,00 & 21,00 &  -3,00 & 17,00 \\
	\bf{1998} && -3,00 & 15,00 &   0,80 & 18,00 & -2,00 & 19,00 &  2,00 & 23,00 &  9,00 & 27,00 & 10,00 & 34,00 & 15,00 & 36,00 & 11,00 & 37,00 & 10,00 & 30,00 &  6,00 & 21,00 & -1,00 & 21,00 &  -5,00 & 13,00 \\
	\bf{1997} && -2,20 & 15,00 &  -2,20 & 15,00 &  0,20 & 20,00 & -1,40 & 20,00 &  4,00 & 27,00 & 11,00 & 30,00 & 11,00 & 33,00 & 12,00 & 33,00 &  9,00 & 32,00 &  3,00 & 26,00 &  2,00 & 18,00 &   1,00 & 16,00 \\
	\bf{1996} && -3,00 & 14,00 &  -4,00 & 13,00 & -6,00 & 15,00 &  1,00 & 21,00 &  6,00 & 26,00 &   -   & 33,00 & 11,00 & 35,00 & 13,00 & 33,00 &  7,00 & 24,00 &  4,00 & 21,00 &  1,00 & 20,00 & -10,00 & 15,00 \\
	\bf{1995} && -4,00 & 15,00 &  -2,60 & 16,00 & -1,20 & 17,00 & -3,00 & 21,00 &  4,00 & 26,00 &  7,00 & 30,00 & 15,00 & 35,00 &  9,00 & 32,00 &  6,00 & 27,00 &  4,00 & 24,00 & -5,00 & 20,00 &  -1,00 & 15,00 \\
	\bf{1994} && -1,00 & 18,00 &  -3,60 & 16,00 &  0,40 & 24,00 & -1,00 & 23,00 &  4,00 & 27,00 & 10,00 & 31,00 & 15,00 & 34,00 & 15,00 & 38,00 &  9,00 & 33,00 &  2,00 & 23,00 &  2,00 & 23,00 &   3,00 & 21,00 \\
	\bf{1993} && -5,00 & 14,00 &  -5,00 & 16,00 & -2,20 & 22,00 &  3,00 & 25,00 &  8,00 & 28,00 & 10,00 & 31,00 & 10,00 & 33,00 & 10,00 & 37,00 &  8,00 & 30,00 &  5,00 & 26,00 & -3,00 & 16,00 &  -2,00 & 17,00 \\
	\bf{1992} && -3,80 & 14,00 &  -5,00 & 16,00 & -1,20 & 20,00 &  2,00 & 23,00 &  5,00 & 28,00 & 10,00 & 29,00 & 12,00 & 35,00 & 14,00 & 36,00 &  8,00 & 30,00 &  5,00 & 25,00 &  2,00 & 20,00 &  -1,80 & 15,00 \\
	\bf{1991} && -5,00 & 13,00 & -10,00 & 19,00 &  0,80 & 25,00 & -1,00 & 19,00 &  1,00 & 23,00 &  8,00 & 30,00 & 10,00 & 34,00 & 12,00 & 33,00 & 11,00 & 28,00 &  3,00 & 24,00 & -3,20 & 17,00 &  -4,00 & 17,00 \\
	\bf{1990} && -4,00 & 15,00 &  -1,60 & 17,00 & -2,60 & 19,00 &  1,00 & 17,00 &  6,00 & 26,00 &  7,00 & 32,00 & 13,00 & 34,00 & 13,00 & 32,00 &  8,00 & 27,00 &  5,00 & 27,00 & -2,20 & 21,00 &  -5,00 & 13,00 \\ \hline
	\bf{Average 1990-2014} && \bf{-3,53} & \bf{15,35} & \bf{-3,24} & \bf{16,06} & \bf{-1,33} & \bf{20,38} & \bf{1,42} & \bf{23,32} & \bf{6,58} & \bf{27,76} & \bf{9,95} & \bf{32,74} & \bf{12,58} & \bf{35,19} & \bf{13,46} & \bf{35,14} & \bf{9,04} & \bf{30,16} & \bf{4,91} & \bf{25,50} & \bf{0,17} & \bf{20,46} & \bf{-2,78} & \bf{16,21} \\
		\hline
	\end{tabular} 
}
\end{table}
\centering
% Table 2 - Summary of Data Meteo-Climatic of Acquapendente \%
\begin{table}
\centering \scriptsize
\caption{Day Time and Night Time Variation} 
\begin{tabular}{|c|c|c|c|c|c|c|c|c|c|c|c|c|}
\hline
\bf Year & \bf Jan &\bf Feb &\bf Mar & \bf Apr &\bf May&\bf Jun & \bf Jul &\bf Aug&\bf Sep&\bf Oct &\bf Nov&\bf Dec \cr
\hline
2014  & 6,32  & 7,21  & 9,48  & 9,27  & 10,03 & 11,27 & 9,52  & 10,97 & 9,43  & 9,61  & 7,53  & 6,03 \\
2013  & 6,90  & 7,39  & 7,35  & 9,33  & 8,61  & 11,03 & 11,84 & 12,26 & 11,07 & 8,26  & 6,10  & 7,61 \\
2012  & 9,24  & 8,26  & 12,26 & 8,87  & 10,13 & 12,55 & 13,35 & 14,10 & 9,03  & 9,42  & 6,73  & 7,13 \\
2011  & 6,87  & 9,32  & 8,13  & 9,93  & 11,65 & 10,93 & 10,68 & 12,88 & 11,81 & 10,19 & 9,63  & 7,74 \\
2010  & 5,48  & 5,68  & 8,87  & 10,16 & 8,20  & 9,74  & 12,50 & 11,97 & 10,00 & 8,61  & 6,74  & 5,37 \\
2009  & 5,68  & 7,86  & 8,42  & 8,07  & 10,16 & 8,03  & 10,00 & 11,26 & 9,47  & 9,45  & 8,57  & 5,45 \\
2008  & 7,52  & 9,70  & 9,26  & 9,85  & 10,43 & 7,93  & 10,39 & 12,61 & 9,80  & 9,39  & 7,53  & 5,48 \\
2007  & 8,78  & 9,54  & 9,64  & 12,77 & 12,53 & 12,01 & 15,22 & 12,77 & 12,18 & 10,45 & 9,01  & 7,88 \\
2006  & 8,54  & 9,25  & 8,72  & 10,68 & 12,81 & 13,82 & 14,38 & 11,43 & 10,97 & 10,50 & 10,04 & 8,18 \\
2005  & 8,21  & 8,71  & 10,90 & 10,84 & 13,37 & 13,80 & 14,37 & 12,55 & 11,04 & 9,30  & 9,76  & 7,75 \\
2004  & 7,40  & 8,31  & 9,23  & 9,73  & 11,83 & 12,94 & 14,67 & 13,62 & 12,42 & 9,37  & 7,33  & 6,50 \\
2003  & 7,51  & 9,81  & 12,58 & 12,55 & 14,76 & 15,25 & 15,68 & 14,91 & 12,29 & 8,60  & 7,65  & 7,25 \\
2002  & 9,82  & 9,34  & 12,19 & 13,16 & 12,02 & 14,09 & 13,16 & 11,34 & 10,21 & 10,46 & 7,07  & 5,52 \\
2001  & 5,93  & 8,66  & 7,78  & 10,99 & 10,61 & 13,77 & 13,57 & 13,92 & 10,90 & 10,95 & 8,51  & 7,65 \\
2000  & 7,94  & 8,37  & 8,23  & 7,77  & 9,93  & 14,69 & 14,29 & 15,50 & 12,13 & 8,72  & 7,10  & 7,30 \\
1999  & 8,47  & 8,24  & 8,68  & 8,60  & 8,94  & 10,03 & 9,74  & 11,61 & 9,43  & 8,10  & 6,91  & 5,63 \\
1998  & 6,91  & 8,73  & 9,29  & 8,45  & 9,23  & 10,68 & 11,13 & 12,20 & 9,29  & 7,77  & 5,57  & 6,75 \\
1997  & 6,88  & 7,75  & 10,86 & 9,39  & 11,96 & 10,29 & 13,42 & 12,45 & 11,45 & 9,10  & 5,84  & 6,09 \\
1996  & 6,50  & 7,82  & 8,90  & 10,10 & 10,45 & 11,93 & 8,43  & 11,55 & 10,03 & 8,43  & 7,30  & 5,30 \\
1995  & 7,64  & 8,31  & 9,75  & 9,79  & 11,45 & 12,80 & 13,32 & 11,10 & 10,37 & 11,06 & 7,91  & 5,65 \\
1994  & 6,54  & 7,11  & 11,97 & 9,18  & 11,03 & 11,03 & 12,77 & 13,68 & 10,39 & 8,55  & 6,19  & 7,23 \\
1993  & 8,18  & 11,83 & 10,57 & 10,57 & 12,61 & 12,93 & 13,73 & 14,24 & 10,10 & 7,28  & 6,21  & 7,88 \\
1992  & 8,59  & 10,37 & 9,52  & 8,97  & 12,16 & 10,20 & 12,48 & 13,61 & 12,57 & 6,97  & 8,00  & 6,81 \\
1991  & 8,50  & 8,44  & 9,36  & 10,47 & 10,48 & 12,07 & 13,61 & 13,87 & 10,87 & 8,71  & 7,53  & 9,30 \\
1990  & 9,24  & 10,36 & 10,42 & 8,52  & 11,24 & 13,44 & 13,81 & 12,87 & 11,10 & 8,13  & 7,93  & 6,20 \\ \hline
\bf Average 1990-2014 & \bf 7,58  & \bf 8,66  & \bf 9,69  & \bf 9,92  & \bf 11,07 & \bf 11,89 & \bf 12,64 & \bf 12,77 & \bf 10,73 & \bf 9,10  & \bf 7,55  & \bf 6,79 \\
\hline
\end{tabular}
\end{table}

% Table of Media Rain
% Table from sheet 'Pioggia Media'

\begin{table}
\tiny
\centering
\caption{Rain/Year [mm]}

\begin{tabular}{|c|c|c|c|c|c|c|c|c|c|c|c|c|c|c|}
\hline
\bf Year  & Jan & Feb & Mar & Apr & May & Jun & Jul & Ago & Sep & Oct & Nov &  Dec & \bf Tot & \bf \%Year \\
\hline
&  &  &  & &  & & & & & & & & &          \cr
\hline
\bf{2014}  & 550,00 & 580,00 & 235,00 & 245,00 & 105,00 & 585,00 & 250,00 & 25,00 & 290,00 & 55,00 & 420,00 & 280,00 & 3620,00 & 9,63\% \\
\bf{2013}  & 420,00 & 220,00 & 250,00 & 190,00 & 215,00 & 240,00 & 405,00 & 30,00 & 215,00 & 415,00 & 220,00 & 50,00 & 2870,00 & 7,64\% \\
\bf{2012}  & - & 135,00 & 20,00 & 115,00 & 155,00 & 5,00 & 5,00 & 15,00 & 285,00 & 255,00 & 120,00 & 100,00 & 1210,00 & 3,22\% \\
\bf{2011}  & 50,00 & 60,00 & 395,00 & 90,00 & 250,00 & 165,00 & 195,00 & 25,00 & 20,00 & 30,00 & 45,00 & 70,00 & 1395,00 & 3,71\% \\
\bf{2010}  & 160,00 & 345,00 & 105,00 & 110,00 & 420,00 & 135,00 & 30,00 & 60,00 & 10,00 & 140,00 & 450,00 & 115,00 & 2080,00 & 5,53\% \\
\bf{2009}  & 41,80 & 82,70 & 63,90 & 55,90 & 17,30 & 203,80 & 68,50 & 46,10 & 75,00 & 35,00 & 55,00 & 95,00 & 840,00 & 2,23\% \\
\bf{2008}  & 69,00 & 50,60 & 95,90 & 104,30 & 108,30 & 98,20 & 35,90 & 5,00 & 128,70 & 60,40 & 161,90 & 235,60 & 1153,80 & 3,07\% \\
\bf{2007}  & 58,30 & 70,30 & 116,20 & 30,30 & 64,40 & 37,10 & 15,00 & 147,00 & 50,10 & 26,10 & 23,30 & 39,10 & 677,20 & 1,80\% \\
\bf{2006}  & 277,30 & 118,70 & 108,50 & 45,80 & 2,50 & 17,90 & 68,00 & 88,10 & 138,90 & 39,30 & 26,10 & 63,10 & 994,20 & 2,65\% \\
\bf{2005}  & 48,40 & 41,20 & 77,80 & 90,70 & 47,50 & 92,80 & 7,60 & 101,30 & 191,70 & 180,40 & 59,30 & 168,00 & 1106,70 & 2,94\% \\
\bf{2004}  & 43,70 & 112,50 & 78,10 & 132,90 & 115,70 & 23,10 & 2,50 & 16,10 & 52,30 & 124,60 & 82,40 & 251,10 & 1035,00 & 2,75\% \\
\bf{2003}  & 86,40 & 11,70 & 23,70 & 30,30 & 83,50 & 133,30 & 14,90 & 42,00 & 104,30 & 80,10 & 134,00 & 44,90 & 789,10 & 2,10\% \\
\bf{2002}  & 32,00 & 37,70 & 15,40 & 40,50 & 65,20 & 43,10 & 88,20 & 132,90 & 138,30 & 21,20 & 82,90 & 123,20 & 820,60 & 2,18\% \\
\bf{2001}  & 142,20 & 44,90 & 84,30 & 76,20 & 35,30 & 17,00 & 15,60 & 5,00 & 31,50 & 11,60 & 62,60 & 14,30 & 540,50 & 1,44\% \\
\bf{2000}  & 30,00 & 25,00 & 105,00 & 150,00 & 145,00 & 62,60 & 61,00 & 82,20 & 26,00 & 127,80 & 108,80 & 86,60 & 1010,00 & 2,69\% \\
\bf{1999}  & 20,00 & 20,00 & 335,00 & 170,00 & 180,00 & 320,00 & 230,00 & 125,00 & 190,00 & 110,00 & 290,00 & 165,00 & 2155,00 & 5,73\% \\
\bf{1998}  & 30,00 & 55,00 & 45,00 & 205,00 & 25,00 & 170,00 & 5,00 & 120,00 & 195,00 & 190,00 & 85,00 & 90,00 & 1215,00 & 3,23\% \\
\bf{1997}  & 270,00 & 30,00 & 35,00 & 165,00 & 10,00 & 60,00 & 25,00 & 40,00 & 15,00 & 175,00 & 205,00 & 65,00 & 1095,00 & 2,91\% \\
\bf{1996}  & 120,00 & 165,00 & 40,00 & 275,00 & 110,00 & 95,00 & 80,00 & 155,00 & 305,00 & 175,00 & 450,00 & 275,00 & 2245,00 & 5,97\% \\
\bf{1995}  & 65,00 & 95,00 & 145,00 & 325,00 & 100,00 & 25,00 & 75,00 & 660,00 & 185,00 & 15,00 & 55,00 & 155,00 & 1900,00 & 5,06\% \\
\bf{1994}  & 45,00 & 75,00 & 5,00 & 345,00 & 35,00 & 210,00 & 95,00 & 15,00 & 70,00 & 110,00 & 130,00 & 155,00 & 1290,00 & 3,43\% \\
\bf{1993}  & 10,00 & 5,00 & 55,00 & 325,00 & 400,00 & 40,00 & - & 54,10 & 150,00 & 125,00 & 215,00 & 230,00 & 1609,10 & 4,28\% \\
\bf{1992}  & 100,00 & 20,00 & 150,00 & 190,00 & 180,00 & 215,00 & 105,00 & 50,00 & 210,00 & 510,00 & 60,00 & 455,00 & 2245,00 & 5,97\% \\
\bf{1991}  & 15,00 & 200,00 & 180,00 & 390,00 & 260,00 & 35,00 & 160,00 & 30,00 & 285,00 & 215,00 & 220,00 & 5,00 & 1995,00 & 5,31\% \\
\bf{1990}  & 45,00 & 65,00 & 60,00 & 405,00 & 95,00 & 50,00 & 125,00 & 145,00 & 60,00 & 270,00 & 235,00 & 140,00 & 1695,00 & 4,51\% \\ \hline
\bf{Average 1990-2014}  & \bf{109,16} & \bf{106,61} & \bf{112,95} & \bf{172,08} & \bf{128,99} & \bf{123,16} & \bf{86,49} & \bf{88,59} & \bf{136,87} & \bf{139,86} & \bf{159,85} & \bf{138,84} & \bf{37586,20} & \\
\hline
\bf &&&&&&&&&&&& \bf Year Average & \bf 1503,45 & \\ \hline
\end{tabular}
\end{table}
\begin{table}
\caption{Wind Average[1990-2014]/Year on Km/h}
\tiny
% Table of Wind Speed
% Table generated by Excel2LaTeX from sheet 'Vento - media'
\begin{tabular}{|c|c|c|c|c|c|c|c|c|c|c|c|c|c|c|c|} 
\hline
\multicolumn{ 14}{|c}{} &            &            \\
\hline
\multicolumn{ 1}{|c|}{Day} &        Jan &        Feb &        Mar &        Apr &        May &        Jun &        Jul &        Ago &        Sep &        Oct &        Nov &        Dec &\bf Average &  \bf Min &  \bf Max  \\
\hline
\multicolumn{ 1}{|c|}{} &            &            &            &            &            &            &            &            &            &            &            &            &            &            &            \\
1     & 23,68 & 21,54 & 19,95 & 21,33 & 15,12 & 18,66 & 16,93 & 18,78 & 21,02 & 14,69 & 18,85 & 22,68 & 19,44 & 14,69 & 23,68 \\
2     & 18,82 & 19,62 & 21,64 & 18,73 & 16,57 & 16,90 & 17,82 & 18,75 & 19,21 & 15,70 & 18,89 & 19,30 & 18,50 & 15,70 & 21,64 \\
3     & 22,05 & 17,88 & 21,00 & 18,08 & 19,80 & 17,23 & 18,13 & 18,78 & 18,15 & 15,77 & 17,57 & 20,89 & 18,78 & 15,77 & 22,05 \\
4     & 22,18 & 19,70 & 22,20 & 20,48 & 19,90 & 17,39 & 17,98 & 19,15 & 19,54 & 18,13 & 19,72 & 17,33 & 19,47 & 17,33 & 22,20 \\
5     & 19,09 & 20,06 & 23,24 & 21,88 & 15,89 & 17,70 & 17,91 & 20,00 & 19,33 & 18,13 & 18,59 & 17,13 & 19,08 & 15,89 & 23,24 \\
6     & 17,73 & 20,44 & 20,05 & 18,38 & 18,02 & 17,25 & 17,76 & 20,21 & 17,19 & 16,42 & 17,30 & 20,83 & 18,46 & 16,42 & 20,83 \\
7     & 15,66 & 18,64 & 21,73 & 20,38 & 17,14 & 17,39 & 19,67 & 17,55 & 18,85 & 19,17 & 21,04 & 21,76 & 19,08 & 15,66 & 21,76 \\
8     & 16,16 & 21,18 & 23,14 & 16,87 & 15,38 & 18,00 & 20,15 & 19,18 & 18,73 & 17,27 & 19,35 & 20,74 & 18,85 & 15,38 & 23,14 \\
9     & 14,48 & 20,68 & 20,81 & 19,89 & 16,36 & 18,43 & 17,54 & 17,73 & 19,77 & 16,98 & 17,72 & 22,89 & 18,61 & 14,48 & 22,89 \\
10    & 14,36 & 20,10 & 21,40 & 21,09 & 15,20 & 17,60 & 17,61 & 20,29 & 17,93 & 16,81 & 19,22 & 22,41 & 18,67 & 14,36 & 22,41 \\
11    & 17,61 & 20,38 & 20,69 & 20,93 & 13,98 & 17,71 & 19,09 & 19,57 & 18,00 & 16,52 & 20,79 & 17,89 & 18,60 & 13,98 & 20,93 \\
12    & 19,24 & 18,70 & 17,52 & 20,80 & 16,68 & 17,25 & 17,91 & 20,42 & 21,09 & 17,17 & 18,48 & 17,70 & 18,58 & 16,68 & 21,09 \\
13    & 17,73 & 18,26 & 18,91 & 19,91 & 16,18 & 16,13 & 17,63 & 17,71 & 18,52 & 17,75 & 18,52 & 19,52 & 18,06 & 16,13 & 19,91 \\
14    & 18,83 & 17,78 & 16,25 & 18,98 & 17,28 & 16,45 & 17,57 & 19,11 & 20,48 & 18,65 & 20,77 & 21,34 & 18,62 & 16,25 & 21,34 \\
15    & 20,85 & 21,73 & 17,20 & 18,46 & 17,36 & 17,70 & 18,91 & 18,17 & 17,89 & 19,08 & 21,45 & 20,98 & 19,15 & 17,20 & 21,73 \\
16    & 18,87 & 23,35 & 19,11 & 20,52 & 16,90 & 16,98 & 18,77 & 17,96 & 18,17 & 18,13 & 25,20 & 19,15 & 19,43 & 16,90 & 25,20 \\
17    & 18,89 & 21,08 & 19,43 & 19,07 & 17,64 & 17,26 & 18,98 & 18,00 & 17,15 & 18,83 & 21,32 & 19,52 & 18,93 & 17,15 & 21,32 \\
18    & 17,59 & 21,40 & 19,75 & 19,35 & 19,30 & 16,82 & 17,87 & 16,71 & 16,95 & 17,26 & 19,48 & 17,57 & 18,34 & 16,71 & 21,40 \\
19    & 18,70 & 20,34 & 20,78 & 18,37 & 20,40 & 17,64 & 18,43 & 17,40 & 16,11 & 18,57 & 18,84 & 17,59 & 18,60 & 16,11 & 20,78 \\
20    & 17,91 & 18,54 & 18,74 & 16,72 & 18,83 & 18,61 & 18,33 & 17,37 & 18,77 & 18,15 & 20,77 & 19,39 & 18,51 & 16,72 & 20,77 \\
21    & 20,16 & 17,06 & 19,57 & 19,57 & 17,52 & 16,70 & 19,13 & 18,72 & 19,13 & 21,54 & 20,09 & 19,26 & 19,04 & 16,70 & 21,54 \\
22    & 20,93 & 18,60 & 19,89 & 17,02 & 16,90 & 19,57 & 18,72 & 17,52 & 15,85 & 17,46 & 17,95 & 19,89 & 18,36 & 15,85 & 20,93 \\
23    & 21,46 & 21,94 & 20,09 & 16,84 & 18,05 & 17,67 & 18,74 & 18,04 & 18,50 & 19,09 & 17,84 & 23,20 & 19,29 & 16,84 & 23,20 \\
24    & 24,02 & 19,58 & 19,46 & 16,43 & 18,23 & 16,83 & 21,80 & 16,80 & 18,08 & 18,13 & 17,29 & 22,50 & 19,10 & 16,43 & 24,02 \\
25    & 24,26 & 18,88 & 21,11 & 17,83 & 17,88 & 18,09 & 18,38 & 17,61 & 19,04 & 17,57 & 16,55 & 22,14 & 19,11 & 16,55 & 24,26 \\
26    & 22,17 & 19,02 & 20,94 & 17,65 & 16,90 & 17,00 & 17,93 & 16,41 & 21,19 & 17,83 & 18,07 & 23,04 & 19,01 & 16,41 & 23,04 \\
27    & 19,04 & 18,75 & 22,42 & 16,98 & 17,08 & 17,42 & 17,87 & 18,50 & 19,08 & 16,67 & 18,05 & 19,78 & 18,47 & 16,67 & 22,42 \\
28    & 17,52 & 22,00 & 20,27 & 14,48 & 16,83 & 18,43 & 18,13 & 19,55 & 17,69 & 17,46 & 18,57 & 21,00 & 18,49 & 14,48 & 22,00 \\
29    & 21,02 & 16,40 & 19,69 & 16,21 & 19,88 & 18,09 & 18,09 & 19,76 & 17,65 & 18,23 & 18,18 & 19,83 & 18,58 & 16,21 & 21,02 \\
30    & 21,48 &       & 20,15 & 15,88 & 17,18 & 17,65 & 18,39 & 20,24 & 17,67 & 18,09 & 19,07 & 18,70 & 18,59 & 15,88 & 21,48 \\
31    & 21,17 &       & 22,44 &       & 16,43 &       & 19,64 & 20,59 &       & 20,50 &       & 21,82 & 20,37 & 16,43 & 22,44 \\
\hline
\bf{Average} & \bf{19,47} & \bf{19,78} & \bf{20,31} & \bf{18,64} & \bf{17,32} & \bf{17,55} & \bf{18,45} & \bf{18,60} & \bf{18,56} & \bf{17,80} & \bf{19,18} & \bf{20,25} & \bf{Average} & \bf{16,06} & \bf{22,09} \\
\bf{Min} & \bf{14,36} & \bf{16,40} & \bf{16,25} & \bf{14,48} & \bf{13,98} & \bf{16,13} & \bf{16,93} & \bf{16,41} & \bf{15,85} & \bf{14,69} & \bf{16,55} & \bf{17,13} & \bf{15,76} &       &  \\
\bf{Max} & \bf{24,26} & \bf{23,35} & \bf{23,24} & \bf{21,88} & \bf{20,40} & \bf{19,57} & \bf{21,80} & \bf{20,59} & \bf{21,19} & \bf{21,54} & \bf{25,20} & \bf{23,20} & \bf{22,19} &       &  \\

\hline
\end{tabular}  
\end{table}
\centering
% Table 5 - Trend of Temeperature of Montefiascone in Jun-Jul-Aug \%
\begin{table}
\centering \scriptsize
\caption{Trend of Temperature of Montefiascone in Jan-Feb-Mar and Jun-Jul-Aug} 
\begin{tabular}{|c|c|c|c|c|c|c|c|c|}
\hline
\bf Year & \bf Jan & \bf Feb &\bf Mar&\bf Jun & \bf Jul &\bf Aug \cr
\hline
\bf{2014} & 18,00 & 19,00 & 21,00 & 24,00 & 19,00 & 17,00 \\
\bf{2013} & 18,00 & 18,00 & 22,00 & 23,00 & 21,00 & 22,00 \\
\bf{2012} & 20,00 & 25,00 & 22,00 & 23,00 & 20,00 & 22,00 \\
\bf{2011} & 20,00 & 20,00 & 23,00 & 19,00 & 23,00 & 22,00 \\
\bf{2010} & 16,00 & 19,00 & 20,00 & 20,00 & 35,00 & 21,00 \\
\bf{2009} & 16,00 & 19,00 & 19,00 & 19,00 & 19,00 & 18,00 \\
\bf{2008} & 21,10 & 6,00  & 20,30 & 20,00 & 18,00 & 19,00 \\
\bf{2007} & 20,50 & 22,40 & 20,40 & 23,10 & 25,10 & 23,30 \\
\bf{2006} & 20,80 & 22,50 & 22,40 & 27,10 & 24,40 & 19,90 \\
\bf{2005} & 18,70 & 19,70 & 27,30 & 31,70 & 26,60 & 22,90 \\
\bf{2004} & 21,50 & 19,50 & 23,00 & 23,70 & 26,30 & 21,90 \\
\bf{2003} & 21,90 & 20,90 & 21,50 & 22,30 & 23,10 & 23,10 \\
\bf{2002} & 22,00 & 17,30 & 26,10 & 26,50 & 22,70 & 22,10 \\
\bf{2001} & 14,50 & 21,10 & 21,10 & 26,30 & 20,90 & 20,50 \\
\bf{2000} & 18,00 & 15,00 & 19,00 & 20,10 & 26,10 & 25,30 \\
\bf{1999} & 21,00 & 19,00 & 20,00 & 14,00 & 18,00 & 20,00 \\
\bf{1998} & 18,00 & 17,20 & 21,00 & 24,00 & 21,00 & 26,00 \\
\bf{1997} & 17,20 & 17,20 & 19,80 & 19,00 & 22,00 & 21,00 \\
\bf{1996} & 17,00 & 17,00 & 21,00 & 33,00 & 24,00 & 20,00 \\
\bf{1995} & 19,00 & 18,60 & 18,20 & 23,00 & 20,00 & 23,00 \\
\bf{1994} & 19,00 & 19,60 & 23,60 & 21,00 & 19,00 & 23,00 \\
\bf{1993} & 19,00 & 21,00 & 24,20 & 21,00 & 23,00 & 27,00 \\
\bf{1992} & 17,80 & 21,00 & 21,20 & 19,00 & 23,00 & 22,00 \\
\bf{1991} & 18,00 & 29,00 & 24,20 & 22,00 & 24,00 & 21,00 \\
\bf{1990} & 19,00 & 18,60 & 21,60 & 25,00 & 21,00 & 19,00 \\ \hline
\bf{Average 1990-2014} & \bf{18,88} & \bf{19,30} & \bf{21,72} & \bf{22,79} & \bf{22,61} & \bf{21,68} \\
\hline

\end{tabular}
\end{table}
\newpage \clearpage
% Table generated by Excel2LaTeX from sheet 'Neve'
\begin{table}
\centering
\caption{Snow in Montefiascone - VT}
\tiny
\begin{tabular}{|c|c|c|c|c|c|c|c|c|c|c|c|c|c|c|c|} 
\hline
\multicolumn{14}{|c}{} &            &            \\
\hline
\multicolumn{1}{|c|}{Years} & \bf Jan & \bf Feb & \bf Mar & \bf Apr & \bf  May & \bf  Jun & \bf  Jul & \bf Ago & \bf Sep & \bf Oct & \bf Nov & \bf Dec &\bf Tot gg 1990-2014 & \bf Average 1990-2014 & \bf Cm  \\
\hline
2014  & - & - & - & - & - & - & - & - & - & - & - & 1 & 1 &  &  25.00 \\
2013  & 1 & 1 & - & - & - & - & - & - & - & - & - & - & 2 &  & 115,00 \\
2012  & - & 2 & - & - & - & - & - & - & - & - & - & - & 2 &  & 115,00 \\
2011  & 1 & - & - & - & - & - & - & - & - & - & - & - & 1 &  &  25,00 \\
2010  & 1 & - & - & - & - & - & - & - & - & - & - & 1 & 2 &  &  45,00 \\
2009  & - & - & - & - & - & - & - & - & - & - & - & - & - &  & -\\
2008  & - & - & 1 & - & - & - & - & - & - & - & - & 1 & 2 &  &   6,10 \\
2007  & - & - & - & - & - & - & - & - & - & - & - & - & - &  & - \\
2006  & - & - & 1 & - & - & - & - & - & - & - & - & - & 1 &  &  36,10 \\
2005  & 3 & - & - & - & - & - & - & - & - & - & - & 1 & 4 &  &  17,00 \\
2004  & - & - & - & - & - & - & - & - & - & - & - & - & - &  & - \\
2003  & - & - & - & - & - & - & - & - & - & - & - & - & - &  & - \\
2002  & 1 & - & 1 & - & - & - & - & - & - & - & - & - & 2 &  &   4,90 \\
2001  & - & - & - & 1 & - & - & - & - & - & - & - & 3 & 4 &  &   4,80 \\
2000  & 1 & - & - & - & - & - & - & - & - & - & - & - & 1 &  &  20,00 \\
1999  & - & - & - & - & - & - & - & - & - & - & 1 & - & 1 &  &  10,00 \\
1998  & - & - & 1 & - & - & - & - & - & - & - & - & - & 1 &  &  20,00 \\
1997  & - & - & - & - & - & - & - & - & - & - & - & - & - &  & - \\
1996  & - & 1 & - & - & - & - & - & - & - & - & - & 3 & 4 &  &  25,00 \\
1995  & 3 & - & - & - & - & - & - & - & - & - & - & - & 3 &  &  25,00 \\
1994  & - & - & - & - & - & - & - & - & - & - & - & - & - &  & - \\
1993  & - & - & - & - & - & - & - & - & - & - & - & - & - &  & - \\
1992  & 2 & - & - & - & - & - & - & - & - & - & - & - & 2 &  &  50,00 \\
1991  & - & 1 & - & 1 & - & - & - & - & - & - & - & - & 2 &  &  40,00 \\
1990  & 1 & - & - & - & - & - & - & - & - & - & - & - & 1 &  &   5,00 \\
\hline
\bf{Average 1990-2014} &\bf{1,56} & \bf{1,25} & \bf{1,00} & \bf{1,00} & \bf{-} & \bf{-} & \bf{-} & \bf{-} & \bf{-} & \bf{-} & \bf{1,00} & \bf{1,67} & \bf{36} & \bf{2,00} & \bf{588,90} \\
\hline
\bf{Years Average} & & & & & & & & & & & & & & \bf{0,08} & \cr  \hline
\end{tabular}
\end{table}%
% Table generated by Excel2LaTeX from sheet 'Riassunto Vento'
\begin{table}[] 
	\centering
	\caption{Average wind speed in the period 1990-2014 for Montefiascone - VT}
	\tiny \rotatebox{90}{
	\begin{tabular}{|lllllllllllllllllllllllllll|} \hline
		\multicolumn{1}{|c|}{Year} & \multicolumn{2}{|c|}{Jan} & \multicolumn{2}{|c|}{Feb} & \multicolumn{2}{|c|}{Mar} & \multicolumn{2}{c}{Apr} & \multicolumn{2}{|c|}{May} & \multicolumn{2}{|c|}{Jun} & \multicolumn{2}{|c|}{Jul} & \multicolumn{2}{|c|}{Aug} & \multicolumn{2}{|c|}{Sep} & \multicolumn{2}{|c|}{Oct} & \multicolumn{2}{|c|}{Nov} & \multicolumn{2}{|c|}{Dec} &        &        \\ \hline
		\multicolumn{1}{|c|}{}                               & Min        & Max        & Min        & Max        & Min        & Max        & Min        & Max        & Min        & Max        & Min        & Max        & Min        & Max        & Min        & Max        & Min        & Max        & Min        & Max        & Min        & Max        & Min        & Max        &        &        \\ \hline
		2014                                               & 7,00       & 46,00      & 8,00       & 50,00      & 8,00       & 43,00      & 6,00       & 37,00      & 8,00       & 43,00      & 11,00      & 37,00      & 8,00       & 35,00      & 8,00       & 37,00      & 8,00       & 100,00     & 7,00       & 41,00      & 7,00       & 43,00      & 5,00       & 48,00      &        &        \\
		2013                                               & 5,00       & 41,00      & 8,00       & 44,00      & 9,00       & 52,00      & 8,00       & 32,00      & 7,00       & 41,00      & 9,00       & 30,00      & 10,00      & 65,00      & 10,00      & 70,00      & 8,00       & 41,00      & 8,00       & 30,00      & 6,00       & 57,00      & 8,00       & 59,00      &        &        \\
		2012                                               & -          & -          & -          & -          & -          & -          & -          & -          & -          & -          & 4,00       & 22,00      & -          & -          & 11,00      & 32,00      & 10,00      & 43,00      & 8,00       & 41,00      & 7,00       & 44,00      & 8,00       & 37,00      &        &        \\
		2011                                               & 3,00       & 31,00      & 5,00       & 36,00      & 5,00       & 33,00      & 4,00       & 35,00      & 4,00       & 25,00      & 7,00       & 22,00      & 7,00       & 37,00      & 10,00      & 41,00      & 8,00       & 39,00      & 8,00       & 46,00      & 9,00       & 9,00       & 9,00       & 9,00       &        &        \\
		2010                                               & 5,00       & 24,00      & 3,00       & 37,00      & 4,00       & 24,00      & 4,00       & 24,00      & 3,00       & 22,00      & 10,00      & 30,00      & 5,00       & 30,00      & 7,00       & 47,00      & 7,00       & 29,00      & 5,00       & 30,00      & 4,00       & 38,00      & 6,00       & 31,00      &        &        \\
		2009                                               & -          & -          & -          & -          & 8,00       & 33,00      & 5,00       & 33,00      & 2,00       & 21,00      & 5,00       & 21,00      & 5,00       & 18,00      & 6,00       & 20,00      & 4,00       & 21,00      & 4,00       & 33,00      & 4,00       & 31,00      & 5,00       & 31,00      &        &        \\
		2008                                               & 5,00       & 55,00      & 6,00       & 55,00      & 8,00       & 42,00      & 15,00      & 47,00      & -          & -          & 5,00       & 19,00      & 14,00      & 14,00      & 10,00      & 23,00      & 6,00       & 6,00       & -          & -          & -          & -          & -          & -          &        &        \\
		2007                                               & 7,00       & 63,00      & 8,00       & 39,00      & 6,00       & 52,00      & 8,00       & 34,00      & 11,00      & 47,00      & 9,00       & 50,00      & 14,00      & 48,00      & 10,00      & 47,00      & 12,00      & 48,00      & 9,00       & 55,00      & 8,00       & 48,00      & 8,00       & 50,00      &        &        \\
		2006                                               & 10,00      & 63,00      & 8,00       & 68,00      & 10,00      & 65,00      & 6,00       & 47,00      & 9,00       & 39,00      & 11,00      & 44,00      & 13,00      & 50,00      & 9,00       & 41,00      & 8,00       & 35,00      & 6,00       & 41,00      & 6,00       & 55,00      & 6,00       & 39,00      &        &        \\
		2005                                               & 1,00       & 59,00      & 9,00       & 61,00      & 7,00       & 41,00      & 11,00      & 47,00      & 8,00       & 41,00      & 10,00      & 48,00      & 15,00      & 44,00      & 11,00      & 42,00      & 8,00       & 44,00      & 7,00       & 35,00      & 7,00       & 42,00      & 4,00       & 54,00      &        &        \\
		2004                                               & 5,00       & 55,00      & 4,00       & 47,00      & 5,00       & 39,00      & 7,00       & 42,00      & 5,00       & 41,00      & 5,00       & 34,00      & 7,00       & 35,00      & 7,00       & 28,00      & 7,00       & 48,00      & 2,00       & 34,00      & 4,00       & 67,00      & 2,00       & 47,00      &        &        \\
		2003                                               & 5,00       & 59,00      & 6,00       & 47,00      & 4,00       & 54,00      & 5,00       & 39,00      & 10,00      & 72,00      & 12,00      & 41,00      & 13,00      & 50,00      & 11,00      & 55,00      & 10,00      & 44,00      & 8,00       & 55,00      & 6,00       & 47,00      & 7,00       & 68,00      &        &        \\
		2002                                               & 4,00       & 72,00      & 8,00       & 76,00      & 8,00       & 79,00      & 6,00       & 54,00      & 8,00       & 54,00      & 8,00       & 43,00      & 6,00       & 42,00      & 7,00       & 35,00      & 6,00       & 34,00      & 5,00       & 37,00      & 5,00       & 57,00      & 3,00       & 39,00      &        &        \\
		2001                                               & 6,00       & 57,00      & 7,00       & 68,00      & 7,00       & 100,00     & 10,00      & 68,00      & 8,00       & 58,00      & 9,00       & 58,00      & 11,00      & 61,00      & 10,00      & 50,00      & 10,00      & 79,00      & 4,00       & 47,00      & 6,00       & 79,00      & 9,00       & 86,00      &        &        \\
		2000                                               & 4,00       & 34,00      & 3,00       & 52,00      & 6,00       & 52,00      & 5,00       & 31,00      & 3,00       & 28,00      & 8,00       & 29,00      & 9,00       & 43,00      & 10,00      & 42,00      & 8,00       & 47,00      & 3,00       & 31,00      & 5,00       & 41,00      & 4,00       & 54,00      &        &        \\
		1999                                               & 5,00       & 55,00      & 6,00       & 54,00      & 7,00       & 55,00      & 8,00       & 37,00      & 8,00       & 26,00      & 10,00      & 28,00      & 5,00       & 41,00      & 9,00       & 34,00      & 7,00       & 37,00      & 4,00       & 35,00      & 7,00       & 61,00      & 6,00       & 59,00      &        &        \\
		1998                                               & 6,00       & 57,00      & 8,00       & 42,00      & 7,00       & 57,00      & 8,00       & 42,00      & 5,00       & 34,00      & 13,00      & 25,00      & 9,00       & 34,00      & 8,00       & 39,00      & 7,00       & 41,00      & 5,00       & 37,00      & 6,00       & 50,00      & 6,00       & 48,00      &        &        \\
		1997                                               & 7,00       & 50,00      & 7,00       & 57,00      & 8,00       & 54,00      & 7,00       & 47,00      & 8,00       & 34,00      & 3,00       & 48,00      & 8,00       & 35,00      & 7,00       & 34,00      & 8,00       & 39,00      & 7,00       & 57,00      & 6,00       & 35,00      & 5,00       & 41,00      &        &        \\
		1996                                               & 6,00       & 41,00      & 7,00       & 52,00      & 4,00       & 54,00      & 4,00       & 42,00      & 6,00       & 42,00      & 7,00       & 52,00      & 7,00       & 52,00      & 8,00       & 34,00      & 7,00       & 42,00      & 7,00       & 37,00      & 9,00       & 57,00      & 8,00       & 65,00      &        &        \\
		1995                                               & 8,00       & 61,00      & 4,00       & 41,00      & 8,00       & 59,00      & 6,00       & 37,00      & 4,00       & 47,00      & 7,00       & 31,00      & 7,00       & 31,00      & 5,00       & 42,00      & 5,00       & 61,00      & 4,00       & 24,00      & 6,00       & 47,00      & 4,00       & 44,00      &        &        \\
		1994                                               & 7,00       & 44,00      & 6,00       & 39,00      & 6,00       & 41,00      & 6,00       & 42,00      & 5,00       & 42,00      & 4,00       & 29,00      & 7,00       & 37,00      & 3,00       & 39,00      & 2,00       & 37,00      & 5,00       & 41,00      & 4,00       & 29,00      & 3,00       & 41,00      &        &        \\
		1993                                               & 5,00       & 55,00      & 7,00       & 50,00      & 5,00       & 55,00      & 6,00       & 46,00      & 5,00       & 18,00      & 7,00       & 18,00      & 7,00       & 31,00      & 7,00       & 39,00      & 6,00       & 63,00      & 8,00       & 39,00      & 4,00       & 42,00      & 5,00       & 47,00      &        &        \\
		1992                                               & 7,00       & 50,00      & 5,00       & 47,00      & 5,00       & 52,00      & 7,00       & 63,00      & 7,00       & 61,00      & 6,00       & 29,00      & 6,00       & 28,00      & 9,00       & 35,00      & 7,00       & 33,00      & 8,00       & 42,00      & 7,00       & 39,00      & 8,00       & 67,00      &        &        \\
		1991                                               & 7,00       & 58,00      & 6,00       & 48,00      & 8,00       & 50,00      & 8,00       & 61,00      & 7,00       & 47,00      & 4,00       & 29,00      & 5,00       & 47,00      & 10,00      & 37,00      & 10,00      & 41,00      & 9,00       & 47,00      & 8,00       & 61,00      & 10,00      & 63,00      &        &        \\
		1990                                               & 1,00       & 42,00      & 1,00       & 61,00      & 1,00       & 42,00      & 4,00       & 41,00      & 2,00       & 41,00      & 3,00       & 29,00      & 3,00       & 41,00      & 4,00       & 42,00      & 2,00       & 41,00      & 4,00       & 42,00      & 10,00      & 72,00      & 7,00       & 61,00      &        &        \\ \hline
		Average 1990-2014                                  & 5,48       & 50,96      & 6,09       & 50,91      & 6,42       & 51,17      & 6,83       & 42,83      & 6,22       & 40,17      & 7,48       & 33,84      & 8,38       & 39,54      & 8,28       & 39,40      & 7,24       & 43,72      & 6,04       & 39,88      & 6,29       & 47,96      & 6,08       & 49,50      &        &        \\
		\hline
		Average Min                                        & 6,74       & 6,74       & 6,74       & 6,74       & 6,74       & 6,74       & 6,74       & 6,74       & 6,74       & 6,74       & 6,74       & 6,74       & 6,74       & 6,74       & 6,74       & 6,74       & 6,74       & 6,74       & 6,74       & 6,74       & 6,74       & 6,74       & 6,74       & 6,74       & 6,74   &        \\
		Average Max                                        & 44,16      & 44,16      & 44,16      & 44,16      & 44,16      & 44,16      & 44,16      & 44,16      & 44,16      & 44,16      & 44,16      & 44,16      & 44,16      & 44,16      & 44,16      & 44,16      & 44,16      & 44,16      & 44,16      & 44,16      & 44,16      & 44,16      & 44,16      & 44,16      & 44,16  &        \\
		\hline
		Average Min                                                & 1,00       &            & 1,00       &            & 1,00       &            & 4,00       &            & 2,00       &            & 3,00       &            & 3,00       &            & 3,00       &            & 2,00       &            & 2,00       &            & 4,00       &            & 2,00       &            & 1,00   & mar-90 \\
		Average Max                                                &            & 72,00      &            & 76,00      &            & 100,00     &            & 68,00      &            & 72,00      &            & 58,00      &            & 65,00      &            & 70,00      &            & 100,00     &            & 57,00      &            & 79,00      &            & 86,00      & 100,00 & mar-01 \\
		\hline
	\end{tabular}
}
\end{table}

% Table generated by Excel2LaTeX from sheet 'Riassunto Vento'
\begin{table}[] 
	\centering
	\caption{$\Delta$ Temperature Average 1990-2014 for Montefiascone - VT}
	\tiny \rotatebox{90}{
		\begin{tabular}{|l|l|l|l|l|l|l|l|l|l|l|l|l|l|l|l|l|l|l|l|l|l|l|l|l|} \hline
			\multicolumn{1}{|c|}{Year} & \multicolumn{2}{|c|}{Jan} & \multicolumn{2}{|c|}{Feb} & \multicolumn{2}{|c|}{Mar} & \multicolumn{2}{c}{Apr} & \multicolumn{2}{|c|}{May} & \multicolumn{2}{|c|}{Jun} & \multicolumn{2}{|c|}{Jul} & \multicolumn{2}{|c|}{Aug} & \multicolumn{2}{|c|}{Sep} & \multicolumn{2}{|c|}{Oct} & \multicolumn{2}{|c|}{Nov} & \multicolumn{2}{|c|}{Dec} \\ \hline
			\multicolumn{1}{|c|}{}                               & Min        & Max        & Min        & Max        & Min        & Max        & Min        & Max        & Min        & Max        & Min        & Max        & Min        & Max        & Min        & Max        & Min        & Max        & Min        & Max        & Min        & Max        & Min        & Max              \\ \hline
\bf{Average 1990-2003} & -3,63 & 15,04 & -3,30 & 16,16 & -1,07 & 20,54 & 0,09  & 22,64 & 6,53  & 27,29 & 9,47  & 32,06 & 12,60 & 34,59 & 12,81 & 35,17 & 8,93  & 28,99 & 4,73  & 24,75 & -0,24 & 19,94 & -2,64 & 15,96 \\

\bf{Average 2003-2014} & -3,41 & 15,74 & -3,16 & 15,94 & -1,66 & 20,19 & 3,13  & 24,18 & 6,64  & 28,36 & 10,56 & 33,62 & 12,56 & 35,96 & 14,27 & 35,09 & 9,18  & 31,65 & 5,15  & 26,46 & 0,70  & 21,12 & -2,96 & 16,53 \\
\bf{Average 1990-2003} & -3,63 & 15,04 & -3,30 & 16,16 & -1,07 & 20,54 & 0,09  & 22,64 & 6,53  & 27,29 & 9,47  & 32,06 & 12,60 & 34,59 & 12,81 & 35,17 & 8,93  & 28,99 & 4,73  & 24,75 & -0,24 & 19,94 & -2,64 & 15,96 \\
\bf{Average 2003-2014} & -3,41 & 15,74 & -3,16 & 15,94 & -1,66 & 20,19 & 3,13  & 24,18 &  6,64  & 28,36 & 10,56 & 33,62 & 12,56 & 35,96 & 14,27 & 35,09 & 9,18  & 31,65 & 5,15  & 26,46 & 0,70  & 21,12 & -2,96 & 16,53 \\
\hline
\bf{$\Delta$ Average 1990-2014} & \bf 0,22  & \bf 0,69  & \bf 0,14  & \bf -0,23 & \bf -0,59 & \bf -0,34 & \bf 3,04  & \bf 1,54  & \bf 0,11  & \bf 1,07  & \bf 1,09  & \bf 1,56  & \bf -0,04 & \bf 1,38  & \bf 1,46  & \bf -0,08 & \bf 0,25  & \bf 2,66  & \bf 0,42  & \bf 1,71  & \bf 0,94  & \bf 1,18  & \bf -0,32 & \bf 0,56 \\
\hline
\end{tabular}
}
\end{table}

% Inser Idro-Geological Map\%
\begin{figure}
\begin{center}
\end{center}
\end{figure}

% Inser Average Day time and Night time variations\%

% Inser Massime Temperature Variations\%
\begin{figure}
\begin{center}
\includegraphics[width=1.3\textwidth{}]{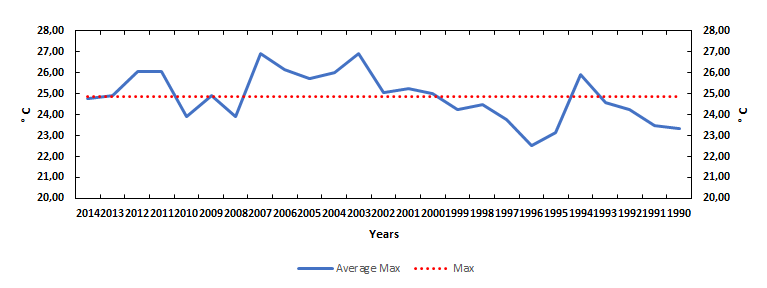}
\caption{Maximum Temperature (dotted line) and Average Temperature Variations 1990-2014 (solid line)}
\end{center}
\end{figure}
% Inser Minime Temperature Variations\%
\begin{figure}
\begin{center}
\includegraphics[width=1.3\textwidth{}]{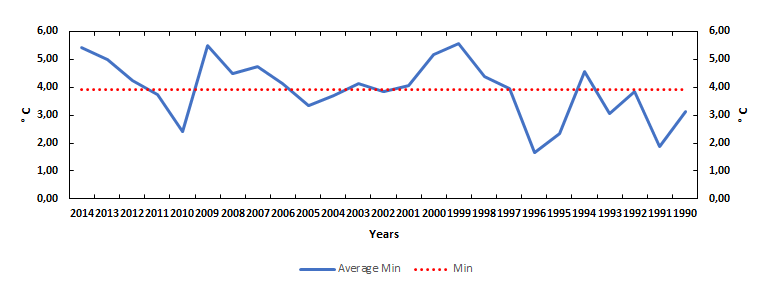}
\caption{Minimum Temperature (dotted line) and Average Temperature Variations 1990-2014 (solid line)}
\end{center}
\end{figure}
% Inser Day of Rain\%
\begin{figure}
\begin{center}
\includegraphics[width=1.3\textwidth{}]{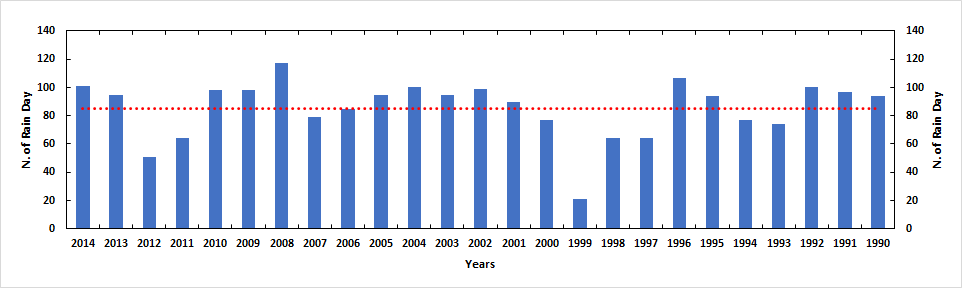}
\caption{Total Days of Rain by Year (solid line) and Average Variations 1990-2014 (dotted line)}
\end{center}
\end{figure}
% Inser Total Rain/YearVariations\%
\begin{figure}
\begin{center}
\includegraphics[width=1.3\textwidth{}]{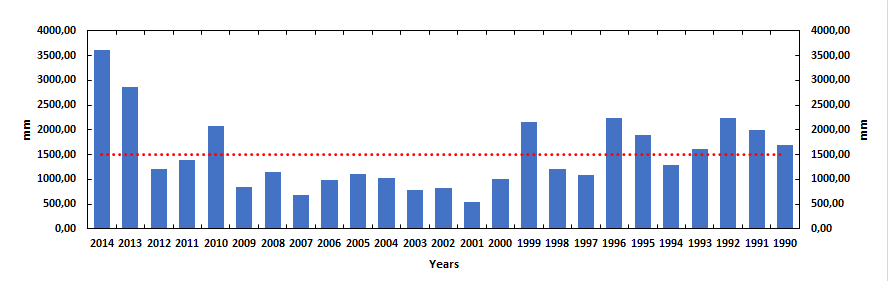}
\caption{Total rain [mm] by year (solid line) and Average Variations 1990-2014 (dotted line)}
\end{center}
\end{figure}
% Inser Day of Fog\%
\begin{figure}
\begin{center}
\includegraphics[width=1.3\textwidth{}]{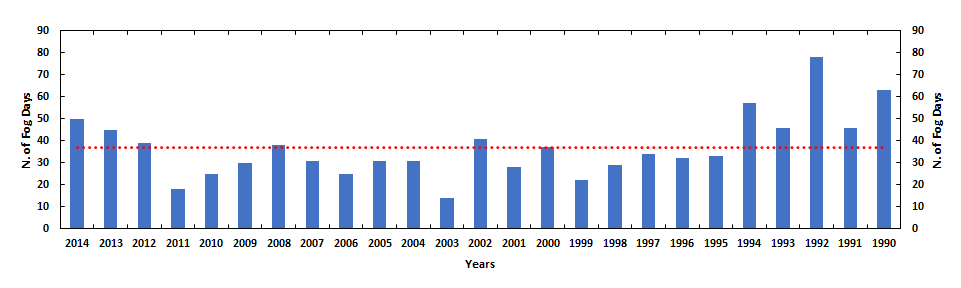}
\caption{Total day of Fog by year (solid line) and Average Variations 1990-2014 (dotted line)}
\end{center}
\end{figure}
% Inser Umidity\%
\begin{figure}
\begin{center}
\includegraphics[width=1.3\textwidth{}]{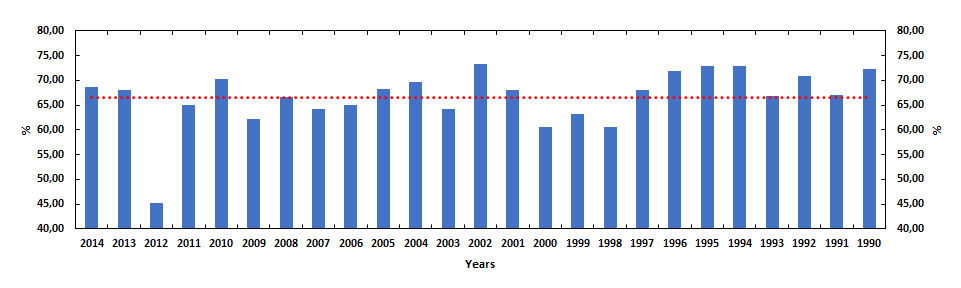}
\caption{Umidity (solid line) and Average Umidity Variations 1990-2014 (dotted line) }
\end{center}
\end{figure}
% Inser Pressure\%
\begin{figure}
\begin{center}
\includegraphics[width=1.3\textwidth{}]{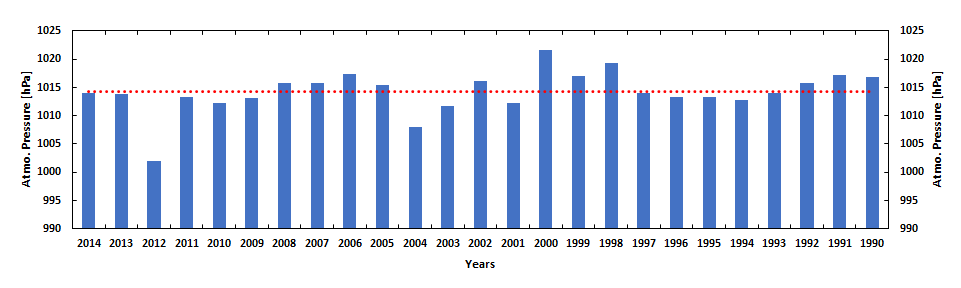}
\caption{Pressure (solid line) and Average Pressure Variations 1990-2014 (dotted line) }
\end{center}
\end{figure}
% Inser Dew_point-Media\%
\begin{figure}
\begin{center}
\includegraphics[width=1.3\textwidth{}]{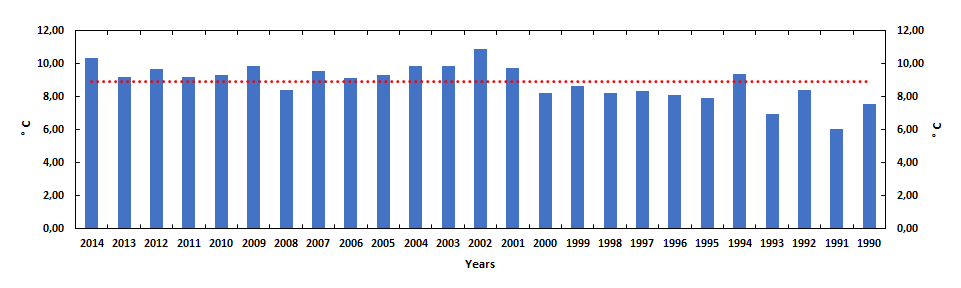}
\caption{Dew Point (solid line) and Average Dew Point Variations 1990-2014 (dotted line) }
\end{center}
\end{figure}
% Inser Wind_Speed_Media\%
\begin{figure}
\begin{center}
\includegraphics[width=1.3\textwidth{}]{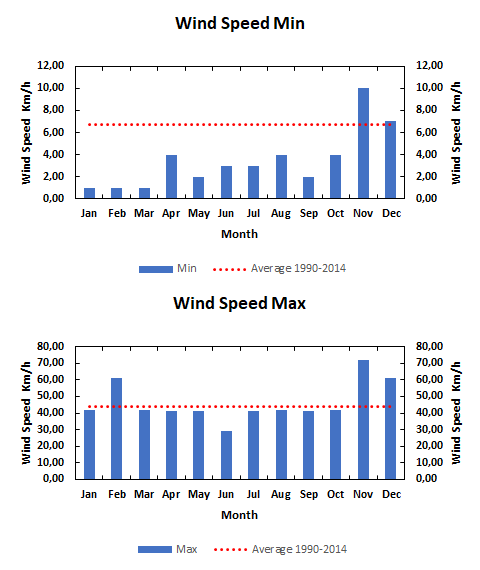}
\caption{Wind Speed (solid line) and Average Wind Speed Variations 1990-2014 (dotted line) }
\end{center}
\end{figure}
\clearpage
% Inser August_Temperature_Media\%
\begin{figure}
\includegraphics[width=1.3\textwidth{}]{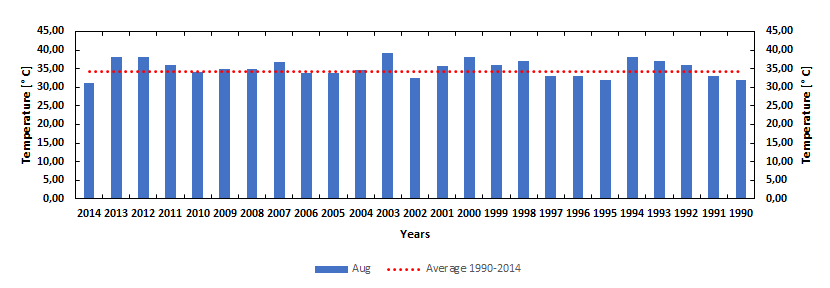}
\caption{August Temperature (solid line) and Average August Temperature 1990-2014 (dotted line) }
\end{figure}
% Inser Febrary_Temperature_Media\%
\begin{figure}
\includegraphics[width=1.3\textwidth{}]{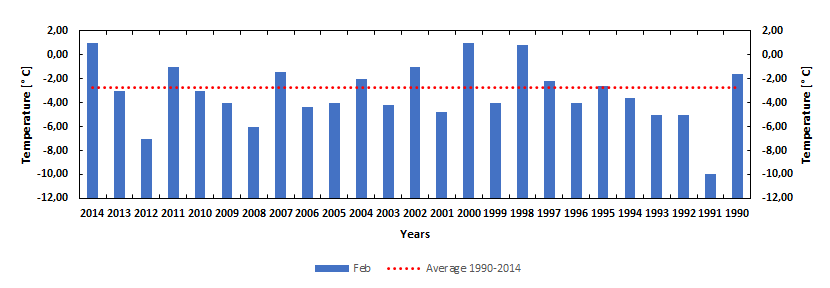}
\caption{February Temperature (solid line) and Average February Temperature 1990-2014 (dotted line)}
\end{figure}

\centering
% Inser Summer_Temperature_Media\%
\begin{figure}
\begin{center}
\includegraphics[width=1.3\textwidth{}]{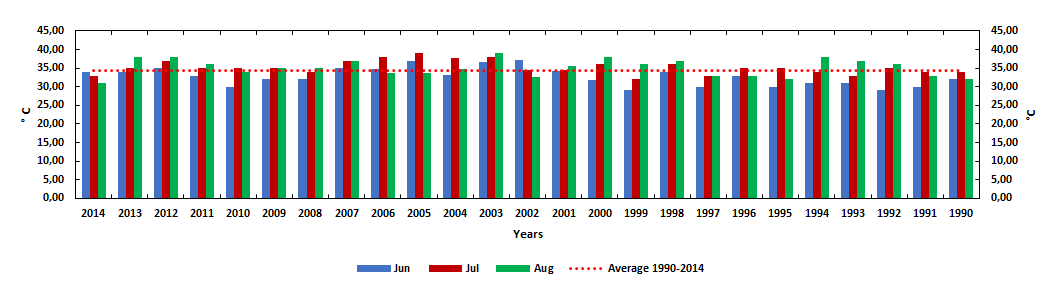}
\caption{Summer Temperature (solid line) and Average Temperature Variations 1990-2014 (dotted line). Jun: blu, Jul: red Aug: green}
\end{center}
\end{figure}
% Inser Winner_Temperature_Media\%
\begin{figure}
\begin{center}
\includegraphics[width=1.3\textwidth{}]{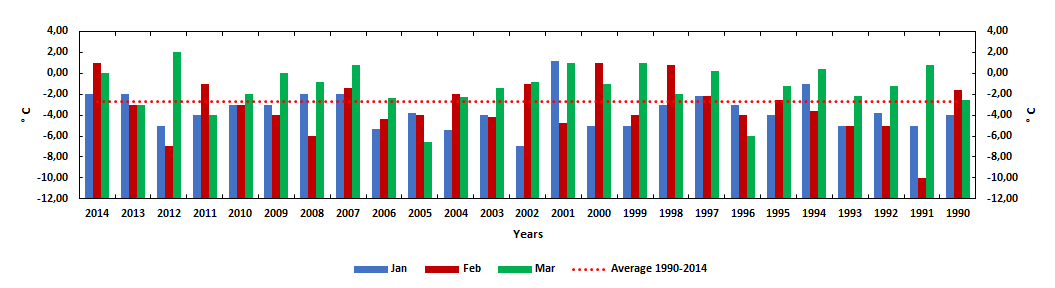}
\caption{Winner Temperature (solid line) and Average Temperature Variations 1990-2014 (dotted line) Jan: blu, Feb: red Mar: green}
\end{center}
\end{figure}
% Inser Total Rain by Month\%
\begin{figure}
\caption{Total Rain (mm) by Month}
\includegraphics[width=0.55\textwidth{}]{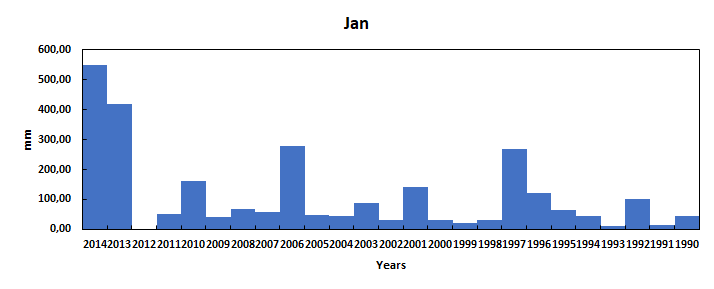} \quad 
\includegraphics[width=0.55\textwidth{}]{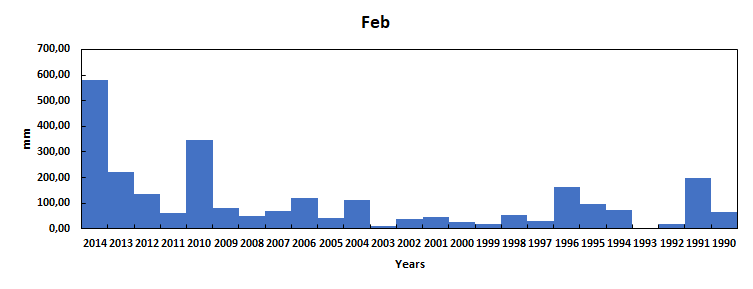} \quad
\includegraphics[width=0.55\textwidth{}]{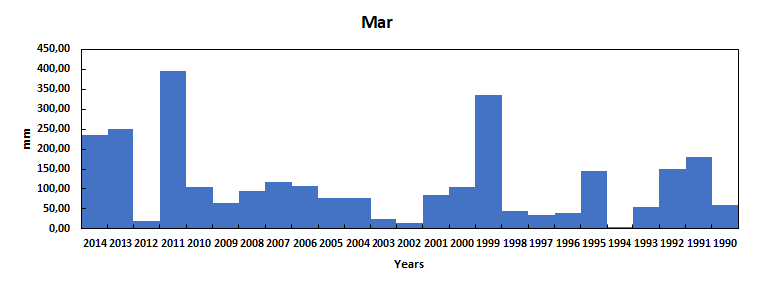} \quad
\includegraphics[width=0.55\textwidth{}]{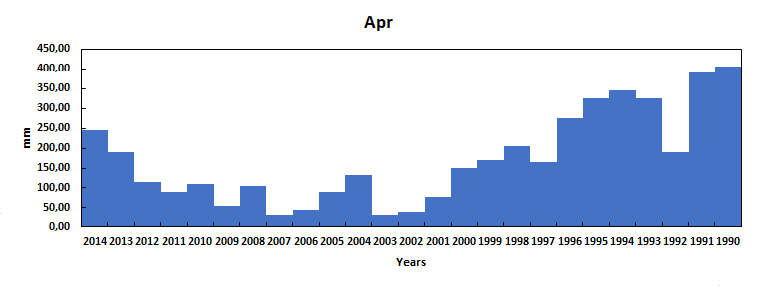} \quad
\includegraphics[width=0.55\textwidth{}]{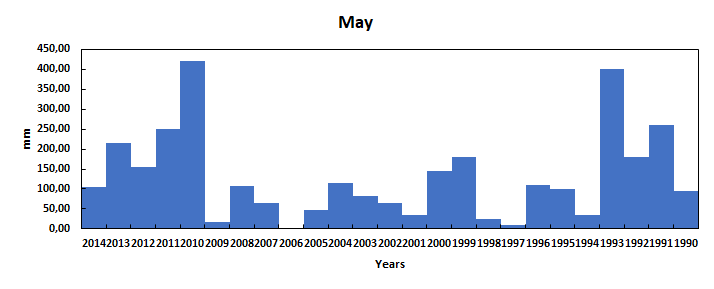} \quad 
\includegraphics[width=0.55\textwidth{}]{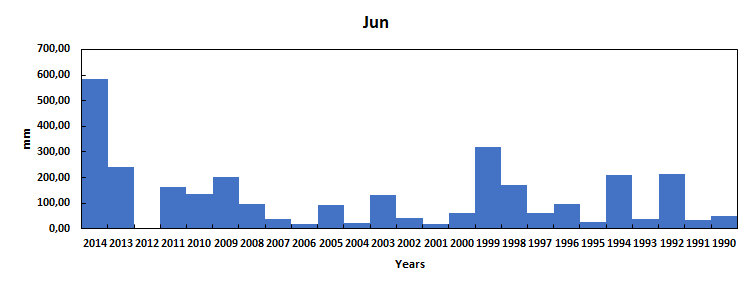} \quad
\includegraphics[width=0.55\textwidth{}]{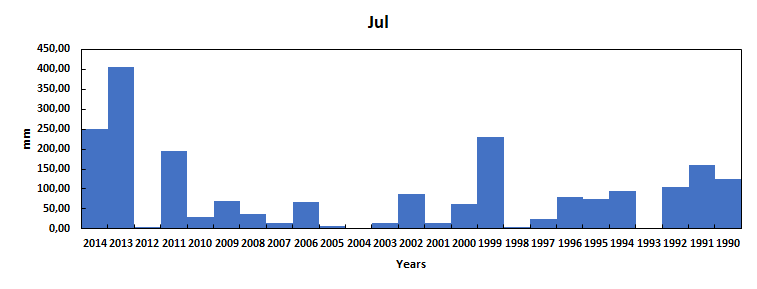} \quad
\includegraphics[width=0.55\textwidth{}]{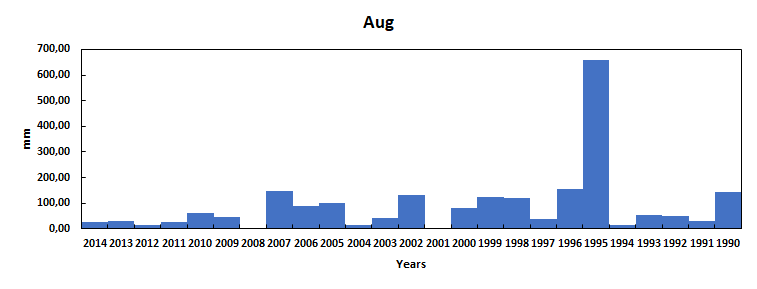} \quad
\includegraphics[width=0.55\textwidth{}]{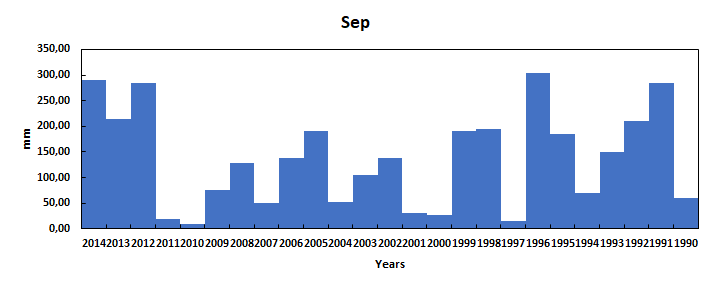} \quad 
\includegraphics[width=0.55\textwidth{}]{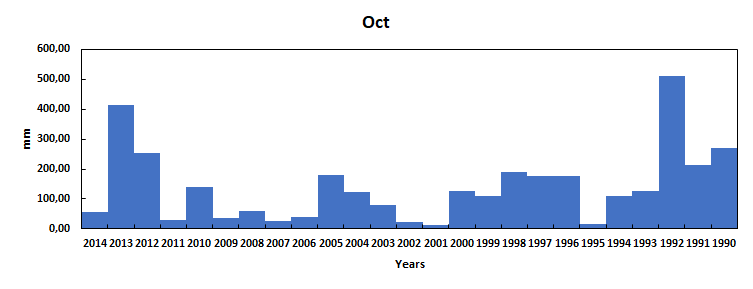} \quad
\includegraphics[width=0.55\textwidth{}]{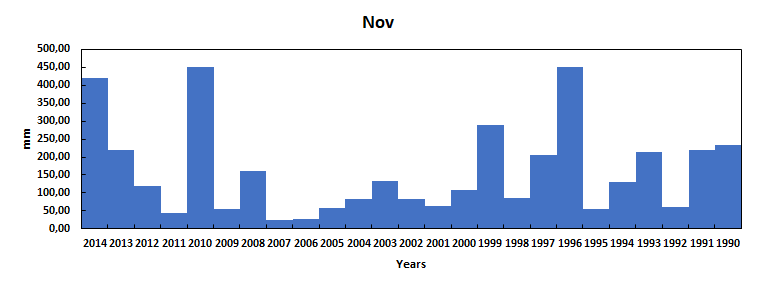} \quad
\includegraphics[width=0.55\textwidth{}]{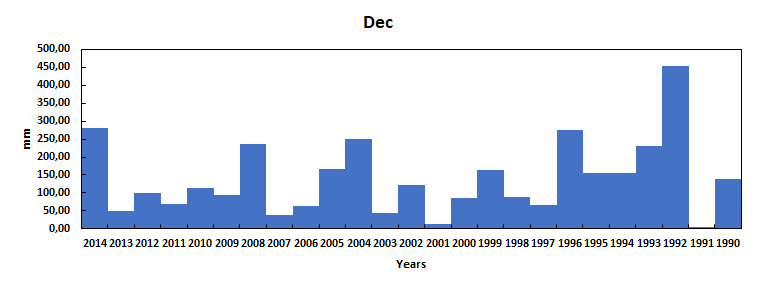} \\
\end{figure}
% Inser Map of Irradiation\%
\begin{figure}
\caption{Solar Irradiation Map}
\includegraphics[width=1\textwidth{}]{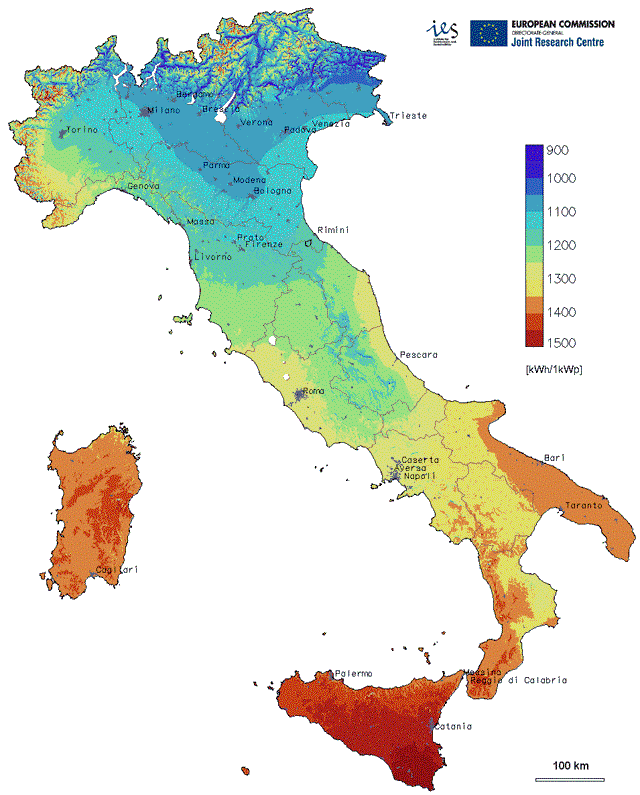}
\end{figure}
% Data of Irradiation\%
\begin{figure}
\includegraphics[width=0.9\textwidth{}]{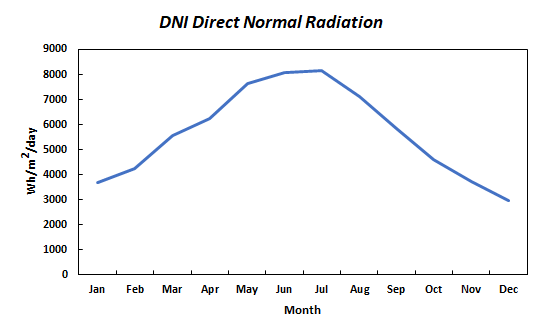}
\caption{DNI Direct Normal Radiation (Wh/m$^2$/day)} 
\end{figure}
% Data of Irradiation\%
\begin{figure}
	\includegraphics[width=0.9\textwidth{}]{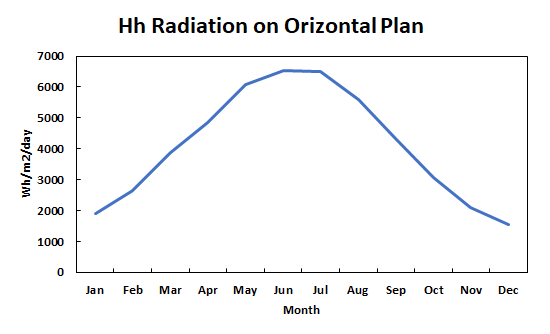}
	\caption{Hh Radiation on Hozizontal Plan (Wh/m$^2$/day)} 
\end{figure}
\begin{figure}
	\includegraphics[width=1.3\textwidth{}]{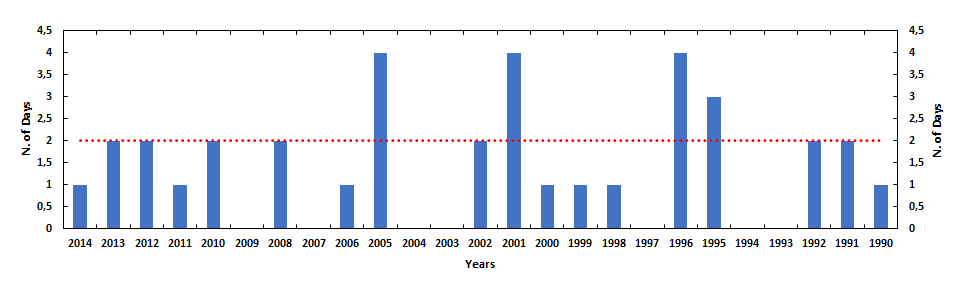}
	\caption{Days of Snow for years} 
\end{figure}
\begin{figure}
	\includegraphics[width=1.3\textwidth{}]{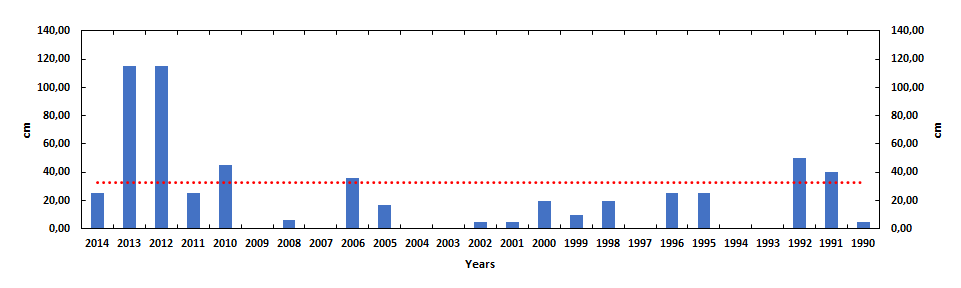}
	\caption{Cm of Snow for years} 
\end{figure}
% Inser Total Wind Month\%
\begin{figure}
	\caption{Total Wind Km/h by Month (solid line) - Average Min (green) and Max (red)}
	\includegraphics[width=0.55\textwidth{}]{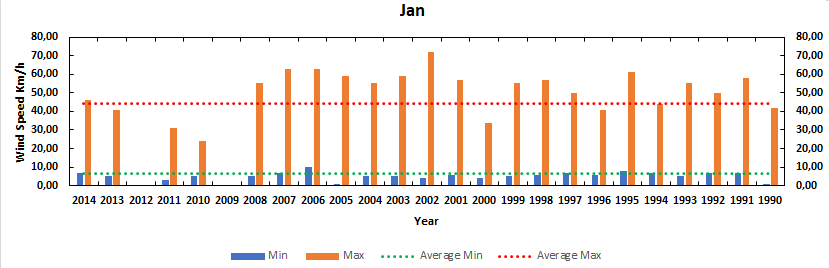} \quad 
	\includegraphics[width=0.55\textwidth{}]{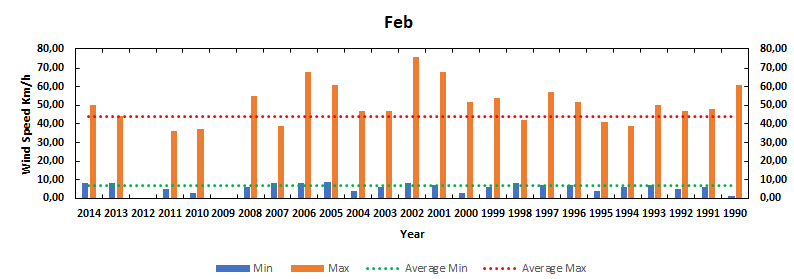} \quad
	\includegraphics[width=0.55\textwidth{}]{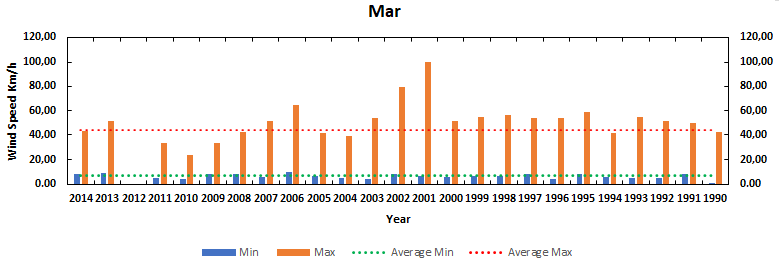} \quad
	\includegraphics[width=0.55\textwidth{}]{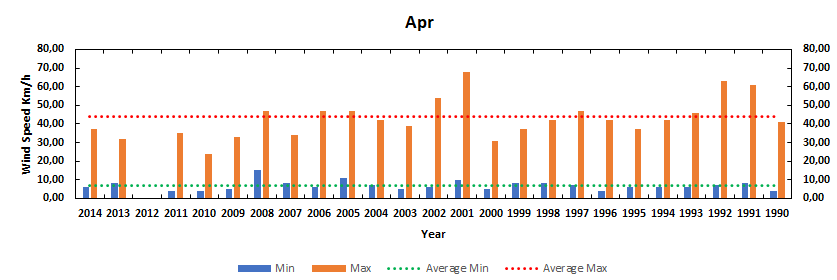} \quad
	\includegraphics[width=0.55\textwidth{}]{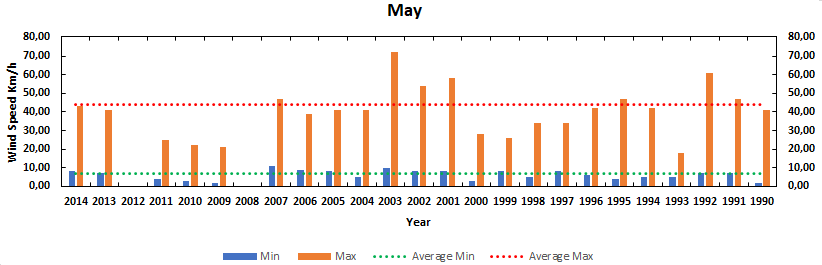} \quad 
	\includegraphics[width=0.55\textwidth{}]{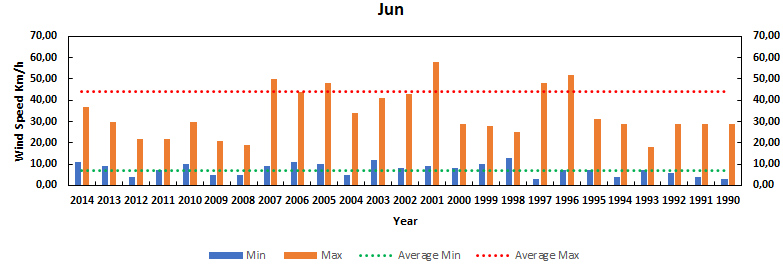} \quad
	\includegraphics[width=0.55\textwidth{}]{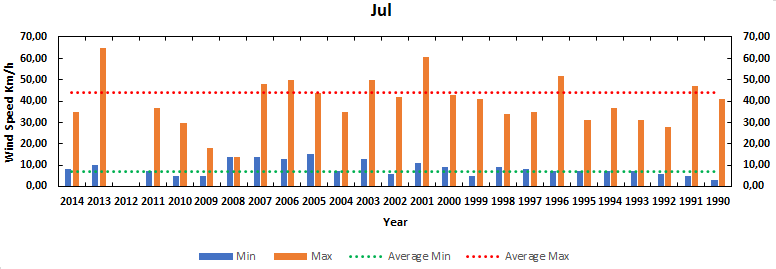} \quad
	\includegraphics[width=0.55\textwidth{}]{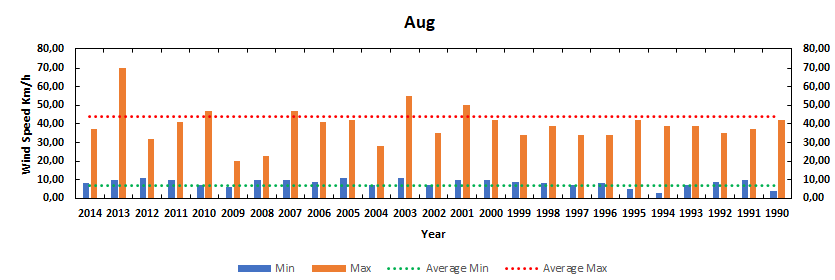} \quad
	\includegraphics[width=0.55\textwidth{}]{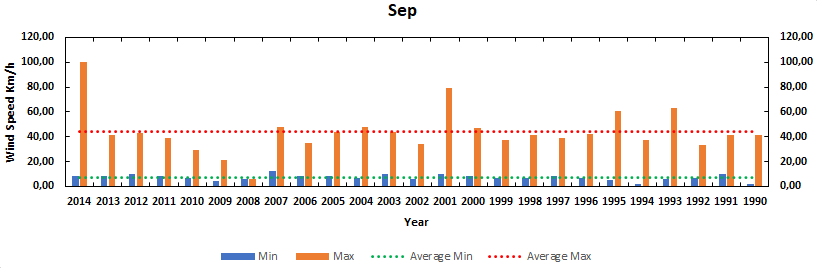} \quad 
	\includegraphics[width=0.55\textwidth{}]{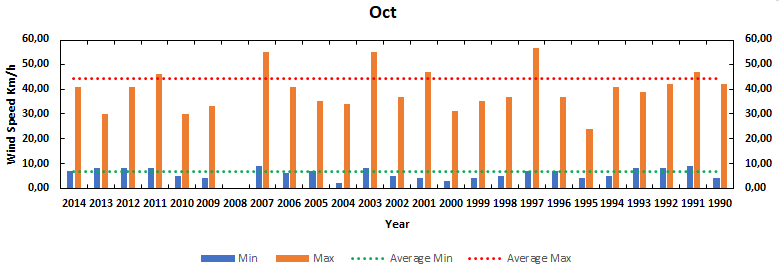} \quad
	\includegraphics[width=0.55\textwidth{}]{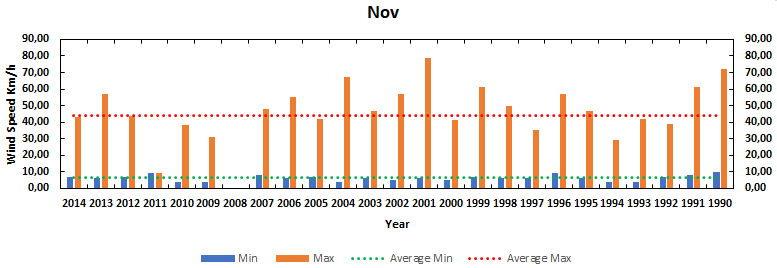} \quad
	\includegraphics[width=0.55\textwidth{}]{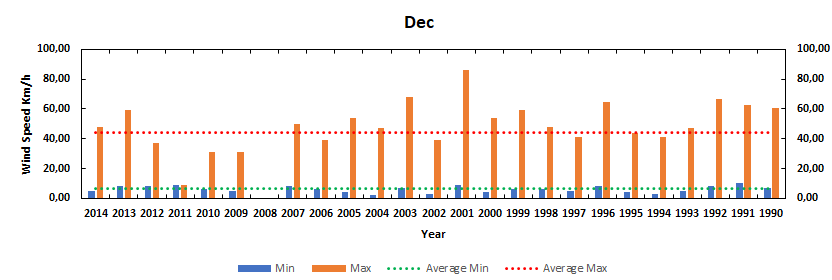} \\
\end{figure}
% Inser Total Wind by Day\%
\begin{figure}
	\caption{Total Wind Km/h by Month (solid line) - Average Min (green) and Max (red)}
	\includegraphics[width=0.55\textwidth{}]{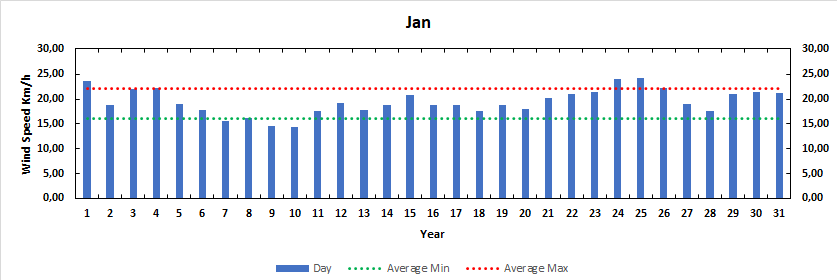} \quad 
	\includegraphics[width=0.55\textwidth{}]{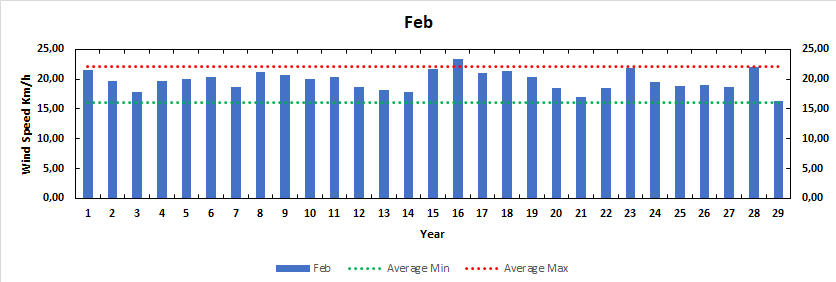} \quad
	\includegraphics[width=0.55\textwidth{}]{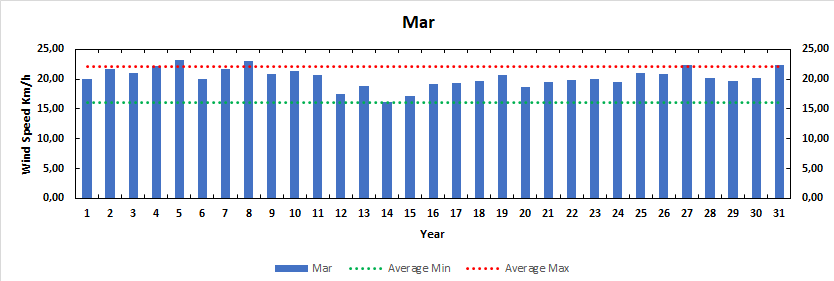} \quad
	\includegraphics[width=0.55\textwidth{}]{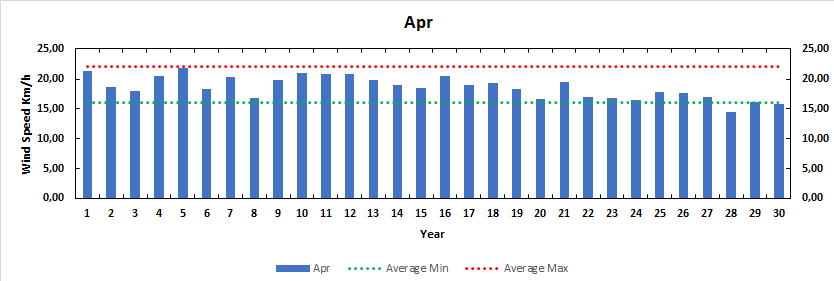} \quad
	\includegraphics[width=0.55\textwidth{}]{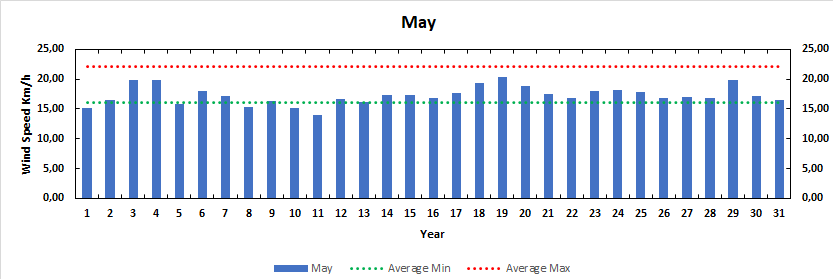} \quad 
	\includegraphics[width=0.55\textwidth{}]{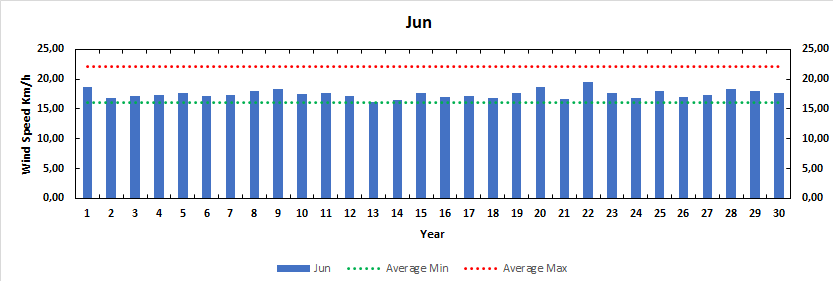} \quad
	\includegraphics[width=0.55\textwidth{}]{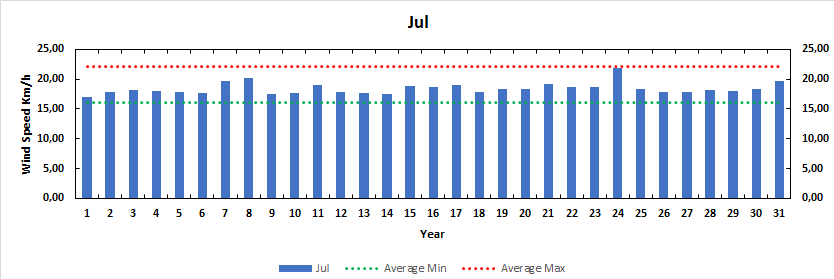} \quad
	\includegraphics[width=0.55\textwidth{}]{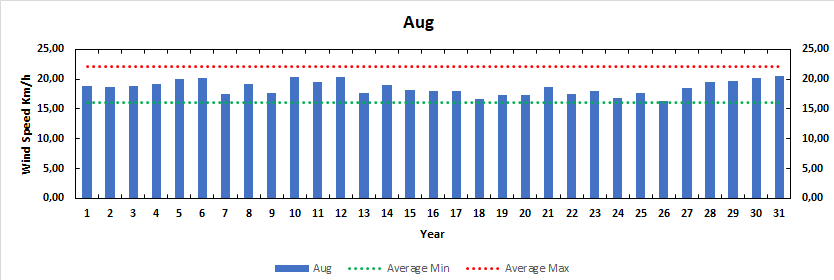} \quad
	\includegraphics[width=0.55\textwidth{}]{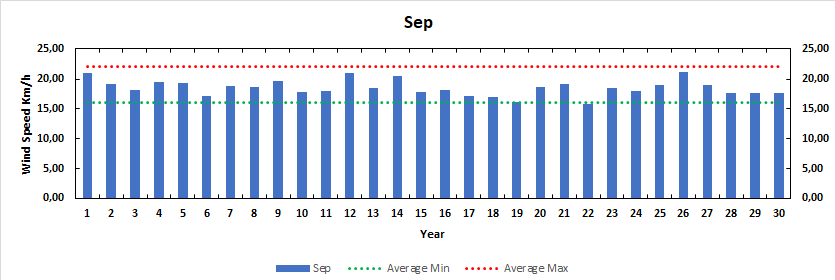} \quad 
	\includegraphics[width=0.55\textwidth{}]{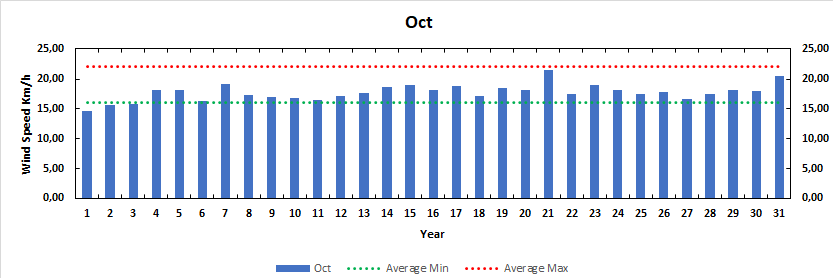} \quad
	\includegraphics[width=0.55\textwidth{}]{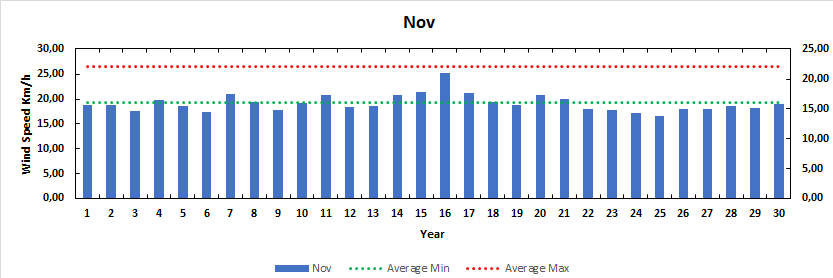} \quad
	\includegraphics[width=0.55\textwidth{}]{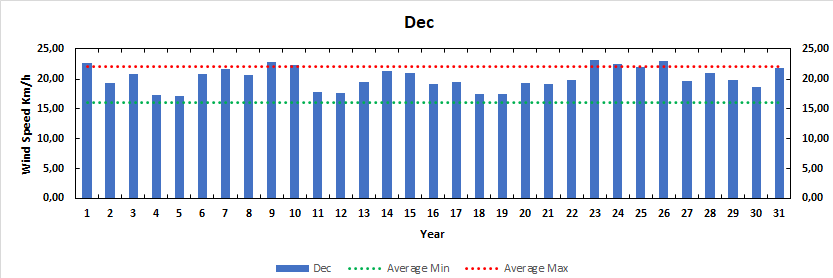} \\
\end{figure}
\clearpage 
\centering
%\phantomsection


\addcontentsline{toc}{section}{\refname}

\begin{thebibliography}{99}
\bibitem{Abbate:1970}
\textbf{Abbate E. \& Sagri M.}, "The eugeosynclinal sequences", Sedimentary Geology, 4: 251 - 340, 1970
\bibitem{Alberti:1970}
\textbf{Alberti A., Bertini M., Del Bono G.L., Nappi G. \& Salvati L.}, "Note Illustrative della Carta Geologica d’Italia alla scala 1:100.000 - Fogli 136 (Tuscania) e 142 (Civitavecchia): pp. 141, 1970 \& 1971
\bibitem{Tasselli:2011ug} 
\textbf {Ente per le Nuove Tecnologie, l'Energia e l'Ambiente}, ``Tabella dei gradi/giorno dei Comuni italiani raggruppati per Regione e Provincia, Legge 26 agosto 1993, n. 412, allegato A'', 1 marzo 2011, p. 151.  
\bibitem{Beccakuva:1991}
\textbf{Beccaluva L., Di Girolamo P. \& Serri G.}, "Petrogenesis and tectonic setting of the Roman Volcanic Province, Italy: evidence for magma mixing", Lithos, 26: 191 - 221, 1991
\bibitem{Buonasorte:1987}
\textbf{Buonasorte G., Fiordelisi A., Pandelli E., Rossi U. \& Sollevanti F.}, "Stratigraphic correlations and structural setting of the pre - neoautochtonous sedimentary sequences of Northern Latium", Period. Mineral., 56: 111 - 122, 1988
\bibitem{Chiocchini:2003}
\textbf{Chiocchini U. \& Madonna S.}, "Geologia delle unità sedimentarie della provincia di Viterbo. Giornata di Studio “Le risorse idriche nel Viterbese: salvaguardia e sviluppo sostenibile", Università della Tuscia, Dipartimento GEMINI. Atti: 7 - 82, 2003
\bibitem{Civitelli:1993}
\textbf{Civitelli G., Corda L.}, "The allochthnous succession of the Sabatini area", note illustrative della carta " Sabatini volcanic Complex”, Quaderni della Ricerca Scientifica 114, V.11 CNR, Roma. 1993
\bibitem{mappageologica:2013} 
\textbf {Prof. L. Carmignani, Prof. G. Cornamusini, Dott. I. Callegari}, ``Mappa Geologica di Montefiascone (VT)'', Università di Siena, Dipartimento di Scienze della Terra - Centro di GeoTecnologie, Foglio 262.
\bibitem{datimeteo:ref2013} 
\textbf {IlMeteo.it} ``Archivio Dati Meteo Montefiascone (VT)'', IlMeteo.it.
\bibitem{eumetsat:ref2013} 
\textbf {Eumetsat} ``Archivio Dati Meteo Montefiascone (VT)'', Eumetsat. 
\bibitem{atlantevento:2013} 
\textbf {Atlantedelvento} ``Archivio Dati Atlante Eolico'', http://www.atlanteeolico.it. 
\bibitem{Peccerillo A:2005}
\textbf {Peccerillo A.} "Plio-Quaternary Volcanism in Italy: Petrology, Geochemistry, Geodynamics, Springer, Heidelberg, 365 pp.", 2005
\bibitem{Dellavedova:2013} 
\textbf {Della Vedova et al.} ``Geothermal structure of Tyrrhenian Sea. Mar. Geol. 55, 271–289'', 1994
\bibitem{Arisi Rota:2013} 
\textbf {Arisi Rota et Fichera} ``Magnetic interpretation connected to Geomagnetic provinces: the Italian case history. 47th Meeting European Association of Exploration Geophysicists Proc.'', 1985
\bibitem{Buonasorte:2013} 
\textbf {Buonasorte et al.} ``Tectonic structures and geometric setting of the Vulsini Volcanic Complex. Per. Mineral., 56: 123-136'', 1987a
\bibitem{Barberi:2013} 
\textbf {F. Barberi, Buonasorte G., R. Cioni, Fiordelisi A., Foresi L., S. Iaccarino, Laurenzi MA , Sbrana A., L. Vernia, Villa IM} ``Plio-Plesitocene geological evolution of the geothermal area of Tuscany and Latium. Mem. Descr. Carta Geol. D’It. XLIX (1994), pp 77-134''
\bibitem{Iaccarino:2013} 
\textbf {Iaccarino et al.} ``Osservazioni stratigrafiche sul bordo orientale del Bacino di Radicofani. Mem. Descr. Carta Geol. D’It. XLIX'', 1994
\bibitem{UE:2013}
\textbf{European Commission} ``Map of Solar Irradiation'', 2012
\bibitem{INGV:2004}
\textbf{INGV} ``Seismic Hazard Map (MPS) of Latium'', 2004
\bibitem{INGV:2004:ref2}
\textbf{INGV} ``S1-INGV Project (http://esse1.mi.ingv.it/ntc.html)'', 2008
\bibitem{Ghelardoni:ref1}
\textbf{Ghelardoni, R.}, "Spostamento dallo spartiacque dell'Appennino Settentrionale in corrispondenza di catture idrografiche", Atti Soc. Sc. Nat., Mem. Ser. A, 65, 1958
\bibitem{Horizon2020:ref1} 
\textbf {Horizon 2020 - Work Programme 2018-2020}, European research infrastructures (including e-Infrastructures), 2016.
\bibitem{Nappi G.:ref1991}
\textbf{Nappi G.}, "Guida all’escursione sui depositi piroclastici del Distretto Vulsino. Fieldtrip Workshop:“Evoluzione dei bacini  neogenici e loro rapporti con il magmatismo plio-quaternario nell’area tosco-laziale”:, pp.45, Pisa 12-13 giugno 1991.
\bibitem{Nappi G.:ref1986}
\textbf{Nappi G., Marini A.}, "I cicli eruttivi dei Vulsini orientali nell’ambito della vulcanotettonica del Complesso". Mem. Soc.Geol. It,35: 679 - 687, 1986
\bibitem{Nappi G.:ref1991-1}
\textbf{Nappi G., Renzulli A., Santi P.}, "Evidence of incremental growt in the Vulsinian calderas (Central  Italy)". Journ.Volc.Geotherm. Res.,47: 13-31, 1991
\bibitem{Nappi G.:ref1994a}
\textbf{Nappi G., Capaccioni B., Mattioli M., Mancini E., Valentini L.}, "Plinian fall deposits from VulsiniVolcanic District (central Italy)", Bull. Volc.,56: 502 - 515, 1994a
\bibitem{Nappi G.:ref1995}
\textbf{Nappi G., Renzulli A., Santi P., Gillot P.Y.}, "Geological evolution and geochronology of the Vulsinian VolcanicDistrict (Central Italy)", Boll. Soc. Geol. It.,114: 599 - 613., 1995
\bibitem{Nappi G.:ref1998}
\textbf{Nappi G., Antonelli A., Coltorti M., Dilani L., Renzulli A., Siena F.}, "Volcanological and petrological evolution of the Eastern  Volcanic District, central Italy", Journ. Volc. Geoth. Res.,87: 211 - 232, 1998
\bibitem{Nappi G.:ref2004}
\textbf{Nappi G., Valentini L., Mattioli M.}, "Ignimbritic deposits in central  Italy: pyroclastic products of the quaternaryage and Etruscan footphats", Field Trip Guide Book  -  P0932° IGC, Florence (Italy), 20 - 28 August 2004: pp. 32, 2004.
\bibitem{Pecceriello A:ref1}
\textbf{Pecceriello A.}, "Plio-Quaternary Volcanism in Italy: Petrology, Geochemistry, Geodynamics, Springer, Heidelberg", 365 pp, 2005
\bibitem{Pecceriello A:ref2}
\textbf{Pecceriello A.}, "Geologia e origini dei vulcani vulsini, Italia centrale", (http://www.orviamm.com/medias/files/peccerillo-geologia-e-origine-dei-vulcani-vulsini.pdf)
\bibitem{Rovida A:ref1}
\textbf{Rovida A., Locati M., Camassi R., Lolli, B., Gasperini P., Antonucci A.}, "Catalogo Parametrico dei Terremoti Italiani (CPTI15), versione 3.0., Istituto Nazionale di Geofisica e Vulcanologia (INGV). https://doi.org/10.13127/CPTI/CPTI15.3, 2021
\bibitem{Stucchi:ref1}
\textbf{Stucchi M., Meletti C., Montaldo V., Crowley H., Calvi G.M., Boschi E.},"Seismic Hazard Assessment
(2003-2009) for the Italian Building Code", Bull. Seismol. Soc. Am. 101(4), 1885-1911, 2011, DOI:
10.1785/0120100130
\end{thebibliography}
\end{document}